%% file: main.tex
\documentclass[journal=jacsat,manuscript=article]{achemso}

\usepackage[version=3]{mhchem} 
\usepackage{float}
\usepackage{amssymb}
\usepackage{amsmath}
\usepackage{algorithm}
\usepackage{verbatim}
\usepackage{cprotect}
\usepackage{algpseudocode}
\usepackage[makeroom]{cancel}
\makeatletter
\newcommand*{\addFileDependency}[1]{
  \typeout{(#1)}
  \@addtofilelist{#1}
  \IfFileExists{#1}{}{\typeout{No file #1.}}
}
\makeatother

\usepackage{hyperref}
\hypersetup{
    colorlinks=true,
    linkcolor=blue,
    citecolor=blue,
    filecolor=magenta,      
    urlcolor=blue,
    pdfpagemode=FullScreen,
    }

\author{Wei-Tse Hsu}
\affiliation{Department of Chemical and Biological Engineering, University of Colorado Boulder, Boulder, CO 80305}
\author{Michael R. Shirts}
\affiliation{Department of Chemical and Biological Engineering, University of Colorado Boulder, Boulder, CO 80305}
\email{michael.shirts@colorado.edu}

\title
  {Replica exchange of expanded ensembles: A generalized ensemble approach with enhanced flexibility and parallelizability}


\begin{document}




\begin{abstract}
Generalized ensemble methods such as Hamiltonian replica exchange (HREX) and expanded ensemble (EE) have been shown effective in free energy calculations for various contexts, given their ability to circumvent free energy barriers via nonphysical pathways defined by states with different modified Hamiltonians. However, both HREX and EE methods come with drawbacks, such as limited flexibility in parameter specification, or the lack of parallelizability for more complicated applications. To address this challenge, we present the method of replica exchange of expanded ensembles (REXEE), which integrates the principles of HREX and EE methods by periodically exchanging coordinates of EE replicas sampling different yet overlapping sets of alchemical states. With the solvation free energy calculation of anthracene and binding free energy calculation of the CB7-10 binding complex, we show that the REXEE method achieves the same level of accuracy in free energy calculations as the HREX and EE methods, while offering enhanced flexibility and parallelizability. Additionally, we examined REXEE simulations with various setups to understand how different exchange frequencies and replica configurations influence the sampling efficiency in the fixed-weight phase and the weight convergence in the weight-updating phase. The REXEE approach can be further extended to support asynchronous parallelization schemes, allowing looser communications between larger numbers of loosely coupled processors such as cloud computing and therefore promising much more scalable and adaptive executions of alchemical free energy calculations. All algorithms for the REXEE method are available in the Python package \verb|ensemble_md|, which offers an interface for REXEE simulation management without modifying the source code in GROMACS.

\end{abstract}


\section{Introduction}
Molecular dynamics (MD) has established its significance and versatility in a broad spectrum of scientific disciplines. With sufficiently accurate force fields, it can theoretically serve as a virtual microscope to investigate a plethora of dynamics of interest at an atomistic resolution. However, even with the latest-generation hardware~\cite{padua2011encyclopedia}, conventional MD simulations are generally limited to only probing processes with relatively short time scales, leaving real-world challenges such as folding of typical proteins or large-scale conformational transitions still out of reach for direct simulation. As such, the last decades have witnessed the emergence of various advanced/enhanced sampling methods dedicated to addressing this timescale issue~\cite{henin2022enhanced}.

These methods can be roughly divided into two broad categories based on different strategies for sampling metastable states of interest that are separated by free energy barriers insurmountable by direct sampling with viable computational cost. Methods in the first category work with the phase space purely defined by the configurational degrees of freedom of the system. Frequently, these methods bias the system along a small number of degrees of freedom (the so-called collective variables, or CVs) to encourage diffusive sampling in the configurational space. Representative examples in this category include umbrella sampling~\cite{umbrella}, metadynamics~\cite{metad}, adaptive biasing force~\cite{ABF}, on-the-fly probability enhanced sampling~\cite{invernizzi2021opes}, and their variations~\cite{var2, invernizzi2022exploration, dama2014transition}. 

\begin{figure}[H]
    \centering
    \includegraphics[width=\textwidth]{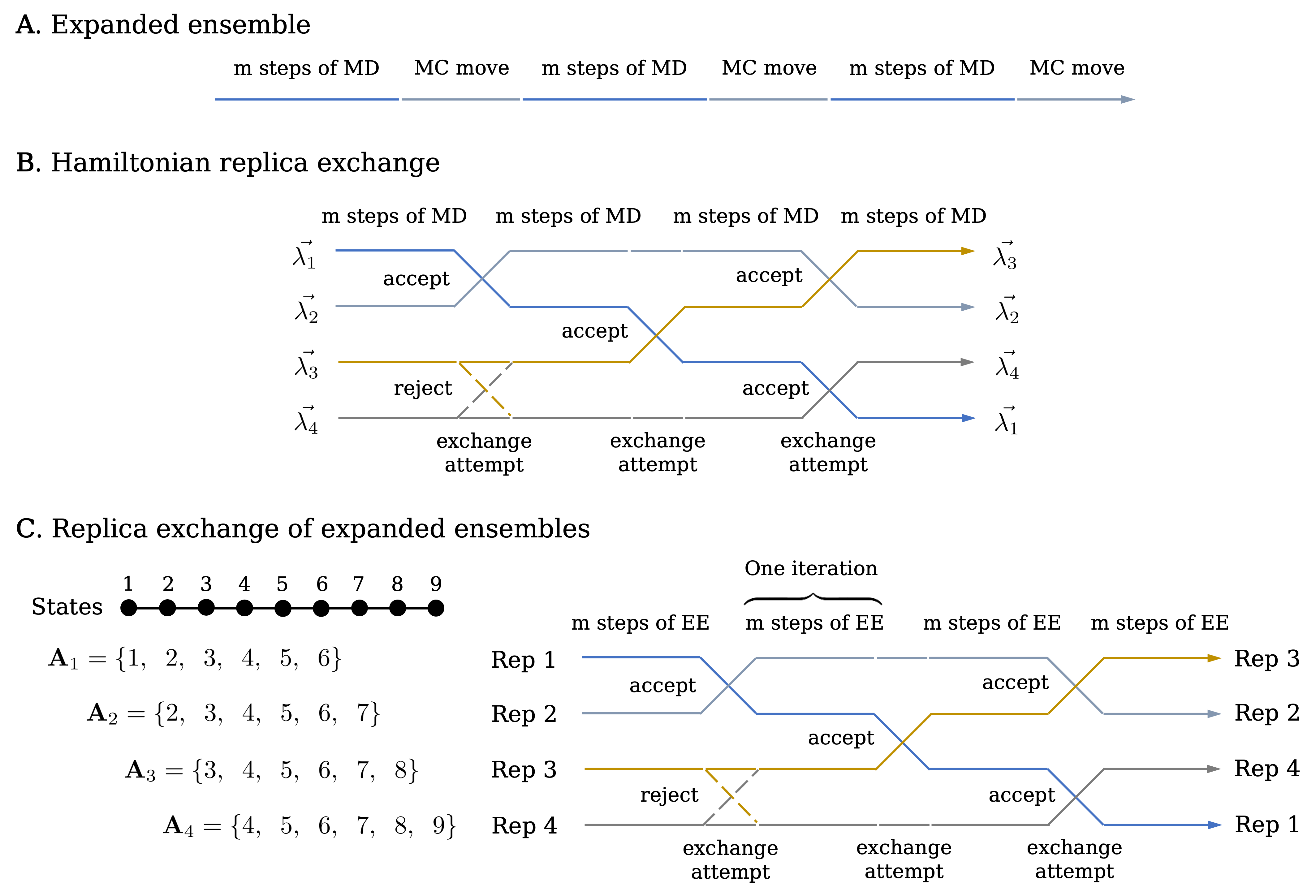}   
    \caption{Schematic representations of different generalized ensemble methods. (A) In \textbf{expanded ensembles}, MD simulations and MC moves alternate periodically to sample the configurational space and the alchemical space, respectively. (B) In \textbf{Hamiltonian replica exchange}, the coordinates of replicas of MD simulations sampling different alchemical intermediate states are periodically exchanged to enhance the sampling in the alchemical space. Each $\lambda$ vector is a vector of coupling parameters of interest, such as for coupling/decoupling electrostatic interactions, van der Waals interactions, or distance restraints. (C) In the \textbf{replica exchange of expanded ensembles} (REXEE) method, the coordinates of replicas of EE simulations are periodically exchanged. ${\bf A}_1$, ${\bf A}_2$, ${\bf A}_3$, and ${\bf A}_4$ denote the sets of states different replicas can sample during the simulation.}
    \label{generalized_ensemble}
\end{figure}

Generalized ensemble methods represent the other category that does not rely on the use of collective variables~\cite{henin2022enhanced}, but expands the phase space by introducing additional dimensions continuously defined or discretized by intermediate states with different temperatures, alchemically modified Hamiltonians, and/or other auxiliary variables. These methods, which can often be expressed as a form of Gibbs sampling~\cite{chodera2011replica, geman1984stochastic}, alternate the sampling direction between the configurational space and the extended space. The sampling in the configurational space is achieved by molecular dynamics (or occasionally Monte Carlo (MC)), while the sampling between different intermediate states in the extended space is usually done by MC moves (panels A and B in Figure \ref{generalized_ensemble}). These moves in the state space use approaches such as the Metropolis-Hastings algorithm~\cite{hastings1970monte}, the Barker transition algorithm~\cite{barker1965monte}, Gibbs sampling~\cite{geman1984stochastic, liu2001monte}, and Metropolized-Gibbs sampling~\cite{liu1996peskun, chodera2011replica}. Importantly, in the case where the metastability of a system changes along the extended state space, sampling the state space in these additional dimensions allows one to move between different metastable states that exist in the configurational space. For example, by traversing the temperature space either serially with a single simulation (e.g., simulated tempering~\cite{marinari1992simulated}) or with multiple simulations running in parallel (e.g., temperature replica exchange (TREX), also known as parallel tempering~\cite{TREMD}), one can observe the unfolded state of a protein that might be unattainable at a lower temperature~\cite{paschek2004reversible, zhou2006replica}. Similarly, alchemical free energy methods exploit the fact that configurational free energy barriers might be lower or even absent in states at intermediate values of the alchemical coupling parameter $\lambda$. With comprehensive sampling in the alchemical space, either serially (e.g., simulated scaling~\cite{li2007simulated}, expanded ensemble (EE)~\cite{EXE}, and $\lambda$ dynamics~\cite{knight2009lambda, knight2011multisite}) or in parallel (e.g., Hamiltonian replica exchange (HREX)~\cite{HREMD}), the system can circumvent free energy barriers via nonphysical pathways bridging the coupled and decoupled states, thus allowing the computation of various free energy differences, including solvation free energies~\cite{mobley2014freesolv, mobley2012alchemical, scheen2020hybrid}, binding free energies~\cite{khalak2021alchemical, lee2020alchemical, abel2017critical, karrenbrock2023nonequilibrium}, and mutation free energies~\cite{piomponi2022molecular, hayes2021strategy}. Importantly, generalized ensemble methods are not confined to solely amplifying the sampling space with the temperature or alchemical space, but have the capacity to define intermediate states and carry out coordinate exchanges in multidimensional grids defined by varying both temperatures and Hamiltonians~\cite{HREMD, bergonzo2014multidimensional, jiang2010free, ebrahimi2019two}, or either of these with other auxiliary variables~\cite{cruzeiro2019multidimensional}. 

To aid the sampling in the extended dimension, especially in methods utilizing alchemical intermediate states, varying forms of weights are usually employed. For instance, expanded ensembles can work with methods such as the Wang-Landau algorithm~\cite{wang2001efficient}, its $1/t$ variation~\cite{belardinelli2007fast, belardinelli2007wang}, accelerated weighted histogram (AWH)~\cite{lidmar2012improving, lindahl2014accelerated, lundborg2021accelerated}, and self-adjusted mixture sampling (SAMS)~\cite{tan2017optimally} to iteratively estimate alchemical weights that aim to level out the alchemical free energy profile. In the limiting case where the set of weights gives each alchemical state exactly equal probability, the alchemical weights are equal to the dimensionless free energies. As another example, in the recently proposed alchemical metadynamics~\cite{hsu2023alchemical} where the alchemical variable is treated as a collective variable, Gaussian biasing potentials deposited in the alchemical direction serve as another form of alchemical weights. 
Hamiltonian replica exchange, while not imposing alchemical weights explicitly, periodically exchanges the coordinates between replicas, which can be regarded as a form of implicit weights enforcing even sampling of each state/replica. 

In expanded ensemble simulations, it is common to adopt a two-stage protocol composed of a weight-updating stage followed by a fixed-weight/production stage. Specifically, a weight-updating algorithm, such as the Wang-Landau algorithm or the similar ones mentioned above, is used in the first stage in to iteratively adjust the alchemical weights until the algorithm is converged, e.g., certain criteria on the flatness of visitation rates at each state are reached. Subsequently, the second stage fixes the weights at the values converged by the first stage. Notably, while data generated in the first stage should theoretically reach quasi-equilibrium given sufficient sampling~\cite{tan2017optimally}, it is of common practice to use samples generated in the second stage for free energy estimators that require equilibrium data, such as thermodynamic integration (TI)~\cite{kirkwood1935statistical}, Bennett acceptance ratio (BAR)~\cite{bennett1976efficient}, and Multistate Bennett acceptance ratio (MBAR)~\cite{shirts2008statistically}. This notion of calculating free energies from equilibrium data also applies to alchemical metadynamics, as discussed in the work~\cite{hsu2023alchemical} by Hsu et al.

Methods such as Hamiltonian replica exchange and expanded ensemble come with limitations. Each simulation within a generalized ensemble must overlap with at least some other states, or no switches can occur. If one is sampling large changes in the phase space, then many connecting states will be necessary, potentially resulting in large numbers of states, with the end-to-end traversal time being diffusion-limited~\cite{escobedo2008optimization} and increasing with the square of the number of states in a given direction.

The selection of the number of states can be potentially problematic for HREX in particular, as running the simulation often needs to take into account the available computational resources, such as the number of CPU cores per node or the total number of nodes, as the simulations must be highly parallelized with relatively low latency to function well. To fully leverage the computational power, the number of replicas should ideally be a factor of the number of available cores. This requirement can become restrictive in deciding the number of states in HREX simulations, especially for complex systems that might necessitate a larger number of states to ensure sufficient overlap between adjacent states. For example, given the one-to-one correspondence between replicas and states, an HREX simulation sampling 64 alchemical states would require 64 replicas. In a situation where a 128-core node is available, but 64 states fail to provide the desired overlap, the next logical increment would be 128 states, which could be an unnecessarily large leap.  These problems can be exacerbated when the system requires hundreds of alchemical intermediate states to ensure adequate overlap between neighboring states, or when the state space is multidimensional. These scenarios can easily require extensive communication between hundreds of cores that current parallelization schemes do not fully support, thereby hindering the exploration of even more complex systems. 

Scenarios that require a large number of states can also pose a challenge to expanded ensembles approaches. Since the sampling along the auxiliary variables is not parallelized, the simulation wall time can therefore be significantly longer than that of the HREX method for equal amounts of sampling at each state.  When weights are poorly converged, as can often happen with adaptive methods when there are slow degrees of freedom orthogonal to the generalized coordinate,~\cite{hsu2023alchemical,belardinelli2007fast, belardinelli2007wang} the round trip time in a 1D-dimensional extended variable space is significantly increased, with potentially some regions not visited at all.  In replica exchange, one may have a lack of round-trip visits along the generalized variable due to insufficient exchanges, but there is at least sampling at all states. For example, it was reported in Refs.~\cite{monroe2014converging} and ~\cite{rizzi2020sampl6} that the Wang-Landau algorithm faced difficulties converging the alchemical weights for several host-guest binding complexes in SAMPL4~\cite{muddana2014sampl4} and SAMPL6~\cite{rizzi2020sampl6} SAMPLing challenges. 

To alleviate some of the aforementioned issues, we propose the method of {\bf replica exchange of expanded ensembles (REXEE)}, which integrates the core principles of the replica exchange (REX) and expanded ensemble (EE) methods. Specifically, a REXEE simulation runs multiple replicas of EE simulations in parallel and periodically exchanges coordinates between replicas. Each EE replica samples a different but overlapping set of alchemical intermediate states to collectively sample the space between the coupled and decoupled states. (See Figure \ref{generalized_ensemble}.) By design, the REXEE method decouples the number of replicas from the number of states, allowing sampling a large number of intermediate states with significantly fewer replicas than those required in the HREX method and other similar methods. By parallelizing replicas, the REXEE method can also reduce the simulation wall time compared to the EE method, and guarantees that there are always simulations, if not at every replica, at least in all ranges of replicas sets along the auxiliary variables. Importantly, such parallelism also sets the stage for wider applications, such as relative free energy calculations for multi-topology transformations. 

We note that there exist replica exchange methods that are well-suited for sampling large numbers of states in a generalized ensemble, with most of them supporting asynchronous parallelization or heterogeneous computational grids. For example, asynchronous replica exchange molecular dynamics~\cite{gallicchio2008asynchronous, gallicchio2015asynchronous, xia2015large, radak2013framework} decentralizes the management of replica simulations by allowing exchanges to be attempted between any two idling replicas. Greedy replica exchange molecular dynamics~\cite{lockhart2015greedy} addressed the synchronization bottleneck in the same spirit, but uses a precomputed schedule of exchange attempts to sample each state equally, as opposed to asynchronous replica exchange. Multiplex replica exchange molecular dynamics~\cite{rhee2003multiplexed} increases the pool of replicas available for exchange attempts, similarly eliminating the need for replica synchronization. We emphasize that these methods are distinctive from the REXEE method as they focus on software implementation design different from the standard replica exchange method. The REXEE method, on the other hand, proposes a new formulation that breaks the one-to-one correspondence between states and replicas, allowing each replica to sample an arbitrary number of states. Still, it can benefit from working with loosely coupled processors with asynchronous parallelization to further enhance the CPU utilization efficiency. 

Upon the development of the REXEE method, this study sets out to accomplish three objectives. First, we seek to demonstrate that with samples from the fixed-weight stage, the REXEE methods can produce free energy estimates on par with those from EE or HREX simulations, while offering greater flexibility in parameter specification and replica configuration. Secondly, by harnessing the statistics gathered from the overlapping states, we examine the ability of the REXEE method to converge alchemical weights compared to the weight-updating EE simulations with the Wang-Landau algorithm~\cite{belardinelli2007fast, belardinelli2007wang}. Importantly, more accurate alchemical weights converged in the weight-updating stage can potentially accelerate the convergence of free energy estimates in the production stage. Lastly, we aim to examine the relationship between REXEE parameters and the performance of the method by comparing various metrics of REXEE simulations with different setups. 

In pursuit of these objectives, we applied the REXEE method with different setups to calculate the solvation free energy of anthracene previously studied by Paliwal et al.~\cite{paliwal2011benchmark}, as well as the binding free energy of the host-guest binding complex CB7-10 from SAMPL4 SAMPLing challenge~\cite{muddana2014sampl4}. With these test systems, we compared the REXEE method with the weight-updating phase of EE simulations in terms of the quality of the converged weights. Additionally, we compared the REXEE method with the HREX simulations and fixed-weight EE simulations regarding the sampling speed and the accuracy of free energy calculations. 

Currently, all necessary algorithms required to enable REXEE have been implemented in the pip-installable Python package \verb|ensemble_md| (\url{https://github.com/wehs7661/ensemble\_md}), which wraps around the key GROMACS functionalities and supports GROMACS~\cite{hess2008gromacs, pronk2013gromacs} starting from version 2022.5. \verb|ensemble_md| serves as an interface to automate and manage the initialization and execution of each iteration in a REXEE simulation. It is equipped with several user-friendly command-line interfaces (CLIs) and has been extensively unit-tested, continuously integrated, and thoroughly documented. Our current implementation of the REXEE method supports only synchronous parallelization, with the development of the asynchronous REXEE method left for future work.

\section{Theory}
\subsection{REXEE configuration}
We consider a REXEE simulation comprised of $R$ parallel replicas of expanded ensembles, each of which is labeled as $i=1, 2, ..., R$, respectively. These $R$ replicas are restricted to sampling $R$ different yet overlapping sets of states (termed  {\bf state sets}) labeled by $m$ as $A_1, A_2, ..., A_R$, which collectively traverse $N$ alchemical intermediate states in total, with $N>R$.  The label $m$ ($m=1, 2, ..., R$) for state sets is a permutation of the label $i$ ($i=1, 2, ..., R)$ for replicas, and vice versa. The relationship between the replica indices and set indices is \begin{equation}
    \begin{cases}
        m = f(i) \\
        i = f^{-1}(m)
    \end{cases}
\end{equation}
where $f$ is a permutation function mapping the replica index $i$ to the state set index $m$ and $f^{-1}$ is its inverse function that performs the reverse mapping. Upon any exchange during the simulation, a new permutation function must be defined. For example, in Figure \ref{generalized_ensemble}C, the first replica is initialized with sampling the first state set, denoted as $f(1)=1$. After an accepted exchange, the first replica samples the second state set in the second iteration, which necessitates a new permutation defined by another function $f'$ such that $f'(1) = 2$ (and $f'(2) = 1$). Importantly, such a permutation relationship implies a one-to-one correspondence between $i$ and $m$, ensuring one and unique state set associated with each replica. 

Additionally, we define $s_i\in \{1, 2, ..., N\}$ as the index of the state currently sampled by the $i$-th replica. For a replica $i$ sampling the state set $A_m$, $s_i$ is additionally constrained such that $s_i \in A_m$. Importantly, the fact that $s_i$ takes values in $\{1, 2, ..., N\}$ and $N>R$ implies a many-to-one relationship between the replica index $i$ and the state index $s_i$, as a certain state may be sampled by multiple replicas. This is in contrast to the one-to-one relationship between the replica index $i$ and the state set index $m$. 

\begin{figure}[H]
    \centering
    \includegraphics[width=\textwidth]{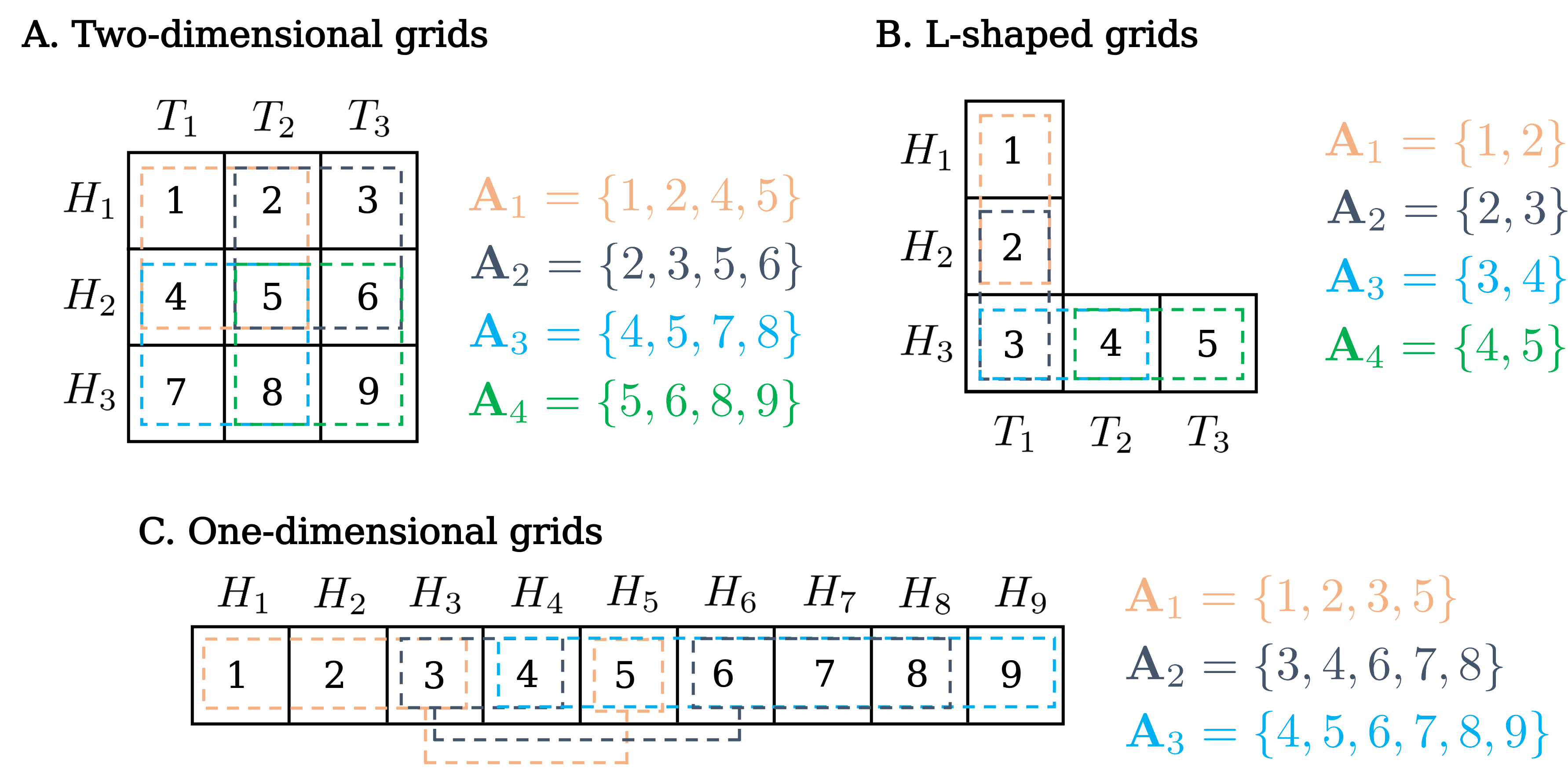}   
    \caption{Different possible replica configurations of a REXEE simulation, with each state represented as a grid labeled by the number in its center and characterized by different Hamiltonians and/or temperatures. Different state sets are represented as dashed lines in different colors. Note that the temperature $T$ and Hamiltonian $H$ can be replaced by other physical variables of interest, such as pressure or chemical potential. (A) Two-dimensional grids, or more specifically, a 3 by 3 grid that defines 9 intermediate states with different temperatures and Hamiltonians. For example, state 8 can be represented as $(T_2, H_3)$. (B) Another 2D example where L-shaped grids define 5 intermediate states with different temperatures and Hamiltonians. (C) One-dimensional grids that define 9 intermediate states with different Hamiltonians.}
    \label{REXEE_more_configs}
\end{figure}

We emphasize that a valid REXEE configuration only requires overlapping state sets and is not restricted to one-dimensional grids, the same number of states for all replicas, nor sequential state indices within the same state sets. For example, Figure \ref{REXEE_more_configs} shows cases where intermediate states are characterized by more than one thermodynamic variable (panels A and B), where different state sets have different numbers of states (panels C), and where the state indices are not consecutive within the same state sets (panels A and C). While some cases presented in Figure \ref{REXEE_more_configs} might not necessarily be practical and are just for illustrative purposes, all configurations in Figure \ref{REXEE_more_configs} fit into the REXEE formalism, as the overlap between different state sets allows the system to access all states upon exchanges. Currently, the most common case is where the intermediate states are defined in a one-dimensional space, with consecutive state indices within the same state set (e.g., the case in Figure \ref{generalized_ensemble}C). In a REXEE simulation adopting such a replica configuration, a state shift $\phi$ between adjacent state sets can be defined indicating to what extent the set of states has shifted along the auxiliary variable, and the simulation can be either homogeneous or heterogeneous, depending on whether all replicas have the same number of states and whether or not the state shift is consistent between all adjacent state sets. 

Given its simplicity and commonality, in this paper, we focus on examples of homogeneous REXEE simulations with one-dimensional alchemical intermediates defined sequentially in each state set, though the approach itself is more general. We denote the number of states per replica as $n_s$, so the $m$-th state set ($m =1, 2, ..., R$) at any given time can be expressed as follows:\begin{equation}A_{m}=\{(m-1)\phi + 1, (m-1)\phi + 2, ..., (m-1)\phi + n_s\}\end{equation} For such a REXEE simulation, the four configurational parameters $N$ (number of states), $R$ (number of replicas), $n_s$ (number of states per state set (or replica)), and $\phi$ (shift between each state set) are related via the following relationship: \begin{equation}
    N = n_s + (R-1)\phi \label{Nnrs}
\end{equation} and we additionally define the overlap ratio $r$ as the ratio between the number of overlapping states and the number of states per replica: \begin{equation}
    r = \frac{n_s - \phi}{n_s}
\end{equation} For example, the configuration of the REXEE simulation shown in Figure \ref{generalized_ensemble}C can be expressed as $(N, R, n_s, \phi)=(9, 4, 6, 1)$ and has an overlap ratio of $r=5/6 \approx 83\%$, as neighboring sets share 5/6 of their states. As discussed in the Supporting Information, solving Equation \ref{Nnrs} with a few additional constraints allows efficient enumeration of all possible REXEE configurations aligned along a single auxiliary variable.

Unlike traditional replica exchange methods, the total number of states $N$ does not have to be equal to the number of replicas $R$ in the REXEE method. In fact, it can be shown that for a REXEE simulation sampling with any number of replicas, there exists at least one valid REXEE configuration (see Figure S1B). This allows a much higher degree of flexibility in the parameter specification as compared to traditional replica exchange methods---once the number of replicas is decided, typically as a factor of the number of available cores, the total number of states can be arbitrary. 

\subsection{State transitions in REXEE simulations}
In a REXEE simulation, regardless of its replica configuration, state transitions occur at both the intra-replica and inter-replica levels. Within each replica of expanded ensemble simulation, transitions between alchemical states within the state set are governed by the selected algorithm in the expanded ensemble simulation, such as the ones mentioned in the section ``Introduction''. The desired probability distribution of each state and detailed balance condition is determined by whichever of various transition schemes implemented for the expanded ensemble method, as discussed in the study by Chodera et al.~\cite{chodera2011replica} Detailed balance for intra-replica exchange ensures the convergence towards an equilibrium distribution for each state set. At the inter-replica level, transitions involve exchanges of configurations between different replicas, which are required to achieve sampling across the entire alchemical space. Detailed balance at this level ensures that the probability influx and outflux are equal for each \textit{set} of states. Notably, these two levels of balance are controlled independently. Both of them need to be obeyed to ensure overall probability distributions and so that the free energy difference across the entire alchemical space can be correctly estimated. 

Since the detailed balance at the intra-replica level can be achieved by simply selecting a well-established method used in traditional generalized ensemble methods~\cite{chodera2011replica}, we only need to derive the acceptance ratio that ensures the detailed balance at the inter-replica level for the REXEE method. Specifically, we consider replicas $i$ and $j$ that sample the state sets $A_m$ and $A_n$, respectively. To swap replicas $i$ and $j$, the state sampled by replica $i$ at the moment, denoted as $s_i \in A_m$, must fall within the state set $A_n$ that is to be swapped, and vice versa. In this case, we call that these replicas $i$ and $j$ are ``swappable'' and we express the exchange of coordinates $x_i$ and $x_j$ between these two replicas as \begin{equation}
    X=\left(..., x^i_{m}, ..., x^j_{n}, ...\right) \rightarrow X' = \left(..., x^j_{m}, ..., x^i_{n}, ...\right)\label{transition}
\end{equation} with $x^i_m\equiv \left(x_i, A_m\right)$ meaning that the $i$-th replica samples the $m$-th state set with the coordinates $x_i$. (See the Supporting information for the strict mathematical definition of the terms involved in Equation \ref{transition}.) Mathematically, the list of swappable pairs $\mathcal{S}$ can be defined as the set of replica pairs as follows:\begin{equation}
    \mathcal{S} = \{(i, j)|s_i \in A_n \text{ and } s_j \in A_m, i\neq j\}\label{swappable}
\end{equation} As discussed in Section 1 in the Supporting Information, the most straightforward way to derive the acceptance ratio is to assume symmetric proposal probabilities, which can be easily achieved by the design of the used proposal scheme. (See the next subsection ``Proposal schemes''.) Under this assumption, the acceptance ratio of swapping the coordinates ($x_i$ and $x_j$) between replicas $i$ and $j$ can be expressed as\begin{equation}
  P_{\text{acc}} = 
  \begin{cases} 
    \begin{aligned}
      &1 &, \text{if } \Delta \leq 0 \\
      \exp(&-\Delta) &, \text{if } \Delta >0
    \end{aligned}
  \end{cases}
  \label{p_acc}
\end{equation} where \begin{equation}
    \Delta = \left(u_{s_i}(x_j) + u_{s_j}(x_i) \right)-\left(u_{s_i}(x_i)+u_{s_j}(x_j)\right)\label{Delta}
\end{equation}
In Equation \ref{Delta}, $u_{s_i}$ and $u_{s_j}$ are the reduced potentials of the states $s_i$ and $s_j$ sampled by replicas $i$ and $j$, respectively. Notably, the expression of $\Delta$ is of the same form as that in the HREX method. This also shows that the REXEE method reduces to the HREX method when each replica is restricted to sampling only one specific state, in which case the state labels $s_i$ and $s_j$ reduce to $i$ and $j$, respectively. (See the Supporting Information for the definition of the reduced potential and the full derivation of the acceptance ratio in the section ``Derivation of the acceptance ratio for swapping EE replicas in a REXEE simulation''.)

\subsection{Proposal schemes}
In this section, we discuss a few common proposal schemes for the REXEE method. Notably, there may easily exist other possible proposal schemes that can achieve the inter-replica level of detailed balance, but we do not further investigate here.

\subsubsection{Single exchange proposal scheme}
The most straightforward proposal scheme is to randomly draw a pair from the list of swappable pairs $\mathcal{S}$ defined in Equation \ref{swappable}, with each pair in the list having an equal probability to be drawn, in which case the proposal probability can be expressed as follows: 
\begin{equation}
\alpha\left(X'|X\right)= \alpha\left(x^j_{m}, x^i_{n} | x^i_{m}, x^j_{n}\right)=
    \begin{cases} 
    \begin{aligned}
      &1/|\mathcal{S}|& \text{, if } (i, j) \in \mathcal{S} \\
      & \quad 0 &\text{, if } (i, j) \notin \mathcal{S}
  \end{aligned}
  \end{cases}\label{p_proposal}
\end{equation}Note that this proposal probability is symmetric, i.e., $\alpha(X'|X)=\alpha(X|X')$ for all $(i, j)$ pairs. With this ``single exchange proposal scheme'', only one exchange is proposed, with the probability defined in Equation \ref{p_proposal}. This proposal scheme has been implemented in the package \verb|ensemble_md| as a sanity check for the REXEE method. 

\subsubsection{Neighbor exchange proposal scheme}
In traditional replica exchange methods, the neighbor exchange proposal scheme alternates between swapping all replica pairs $2\ell-1$ with $2\ell$ and all pairs $2\ell$ with $2\ell+1$ for each $\ell=1, ..., \lfloor R/2\rfloor$. While this notion of alternating between ``odd-even pairs'' and ``even-odd pairs'' is applicable to the REXEE method, one needs to take into account the fact that some pairs may not be swappable, i.e., $(2l-1, 2l)\notin \mathcal{S}$ or $(2\ell, 2\ell+1)\notin \mathcal{S}$ for specific $\ell$ values, with $\mathcal{S}$ defined in Equation \ref{swappable}.

Another way to perform neighbor exchanges, which has been implemented in the Python package \verb|ensemble_md|, is to add a constraint to $\mathcal{S}$ defined in Equation \ref{swappable} such that the swappable pairs consist exclusively of neighboring replicas, with each pair having an equal probability to be drawn. Formally, the proposal probability in this case can be expressed as follows:
\begin{equation}
\alpha\left(X'|X\right)= \alpha\left(x^j_{m}, x^i_{n} | x^i_{m}, x^j_{n}\right)=
    \begin{cases} 
    \begin{aligned}
      &1/|\mathcal{S}_{\text{neighbor}}|& \text{, if } (i, j) \in \mathcal{S_{\text{neighbor}}} \\
      & \quad 0 &\text{, if } (i, j) \notin \mathcal{S_{\text{neighbor}}}
  \end{aligned}
  \end{cases}
\end{equation} where 
\begin{equation}
    \mathcal{S}_{\text{neighbor}} = \{(i, j)|s_i \in A_n \text{ and } s_j \in A_m \text{ and } |i-j|=1\}
\end{equation}
Similarly, the proposal probability in this case is also symmetric. This proposal scheme has also been implemented in the package \verb|ensemble_md| as a sanity check for the REXEE method. 

\subsubsection{Multiple exchange proposal scheme}
As opposed to the single exchange or neighbor exchange proposal schemes, one can propose multiple swaps within an exchange interval to further enhance the mixing of replicas. Importantly, whether two replicas are swappable not only depends on the state sets of the two replicas, but also on the states being sampled by the two replicas at the moment. Therefore, it is not always feasible to propose multiple swaps all at once and perform them serially, as a swappable pair might become unswappable after a previous swap is accepted. To address this issue, the ``multiple exchange proposal scheme'' (see Algorithm \ref{multiple}) proposes swaps one at a time, and whenever a proposed swap is accepted, it updates the permutation of the state sets and re-identifies the list of swappable pairs before proposing the next swap. Note, however, since the order of swapping could influence the resulting swapped configurations when there is any replica involved in multiple proposed swaps, this proposal scheme does not have a symmetric proposal probability. Accordingly, an acceptance ratio other than Equation \ref{Delta} has to be carefully designed to deal with this asymmetry so that the detailed balance at the inter-replica level is obeyed. Currently, this proposal scheme has not been implemented in \verb|ensemble_md| given our focus on proposal schemes with symmetric proposal probabilities.
\begin{algorithm}[H]
\caption{Multiple exchange proposal scheme}
\begin{algorithmic}[1]
\State Identify the list of swappable pairs $\mathcal{S}$.
\For{$n = 1$ to $n_{\text{ex}}$, where $n_{\text{ex}}$ is the desired number of swaps}
\If{$\mathcal{S} \neq \emptyset$}
\State Draw $(i, j)\in \mathcal{S}$ with the proposal probability defined in Equation \ref{p_proposal}.
\State Calculate the acceptance ratio $P_\text{acc}$ for the drawn pair using Equation \ref{p_acc}.
\If{the proposed swap $(i, j)$ is accepted}
\State Perform the swap $(i, j)$, then update the list of swappable pairs $\mathcal{S}$.
\EndIf
\Else
\State \textbf{break}
\EndIf
\EndFor
\end{algorithmic}\label{multiple}
\end{algorithm}

\subsubsection{Exhaustive exchange proposal scheme}
Another approach to carry out multiple swaps serially within an exchange interval is the ``exhaustive exchange proposal scheme'', as detailed in Algorithm \ref{exhaustive}. In brief, it operates similarly to the single exchange proposal scheme, but exhaustively traverses the list of swappable pairs while updating the list by eliminating any pair involving replicas that appeared in the previously proposed pair. This elimination process circumvents the issue of the result depending on the order of swapping, as no replica will be involved in more than one swap and all the swaps proposed in the same exchange interval are independent of each other. Consequently, this ensures that the proposal probability is symmetric and the detailed balance condition is obeyed with the use of Equation \ref{Delta}.  This method has also been implemented in the Python package \verb|ensemble_md|.

\begin{algorithm}[H]
\caption{Exhaustive exchange proposal scheme}
\begin{algorithmic}[1]
\State Identify the list of swappable pairs $\mathcal{S}$.
\While{$\mathcal{S}\neq\emptyset$}
\State Draw $(i, j)\in \mathcal{S}$ with the proposal probability defined in Equation \ref{p_proposal}.
\State Calculate the acceptance ratio $p_\text{acc}$ for the drawn pair using Equation \ref{p_acc}.
\If{the proposed swap $(i, j)$ is rejected}
\State \textbf{break}
\EndIf
\State Perform the swap and update $\mathcal{S}$ by removing pair(s) involving replicas $i$ and $j$.
\EndWhile
\end{algorithmic}\label{exhaustive}
\end{algorithm}

\subsection{Weight combination schemes for weight-updating REXEE}
One pronounced difference between the REXEE approach and other generalized ensemble methods like EE or HREX lies in the existence of ``overlapping states'' in REXEE. These states fall within the intersection of at least two state sets (see Figure \ref{generalized_ensemble}C) and are therefore accessible by multiple simulation replicas. To leverage the samples of these overlapping states collected from multiple replicas in weight-updating REXEE simulations, in a weight-updating REXEE simulation, we could combine alchemical weights for these states across replicas before initializing the next iteration. The hypothesis is that such on-the-fly modifications to the weights could potentially further accelerate the convergence of the alchemical weights during the weight-updating phase, providing a better starting point for the subsequent production phase. 

While there are various possible ways to combine weights across replicas, some of them suffer from reference-dependent results, or the issue of interdependence in the weight difference between adjacent states. For example, one intuitive way to combine weights for a state $s$ across replicas is to calculate the negative logarithm of the probability averaged over all replicas accessible to $s$. Denoting these replicas as $k \in \mathcal{Q}_s$, the combined/averaged weight for state $s$, $\overline{g_{s}}$ (with its corresponding probability $\overline{p_{s}}$), can be expressed as:
\begin{equation}
    \overline{g_{s}} = -\ln \overline{p_{s}} = -\ln \left(\dfrac{1}{|\mathcal{Q}_s|} \sum_{k\in \mathcal{Q}_s} p^k_{s}\right) = -\ln \left (\dfrac{1}{|\mathcal{Q}_s|} \sum_{k\in \mathcal{Q}_{s}} e^{-g^{k}_{s}}\right)
\end{equation}
where $g^k_{s}$ is the alchemical weight of state $s$ in the state set $k$, and $p^k_{s}$ is its corresponding probability. However, as free energy differences are only defined up to a constant, it is standard to set the weight of some reference state to 0. Different replicas sample different states and thus must generally define different reference states. As such, different choices of references could lead to different resulting combined weights and there is no justification about which reference should be favored. 

An alternative approach is to take advantage of the weight differences between adjacent states. Given the weight differences between neighboring states for each replica, this method then calculates the average for the weight differences accessible by multiple replicas.  This average can be either a simple average or the average weighted by the inverse statistical variance of the alchemical weights combined. In the latter case, measurements having lower variability are assigned with higher contributions, so the resulting weighted mean is less sensitive to outliers. In this study, we write the weight difference between the states $s$ and $s+1$ in replica $i$ as $\Delta g_{(s, s+1)}^i=g^i_{s+1}-g^i_{s}$, and the set of replicas that can access both $s$ and $s+1$ as $\mathcal{Q}_{(s, s+1)}$. Then, for the case where the inverse-variance weighting is used, we have the averaged weight difference between $s$ and $s+1$ as: 
\begin{equation}
    \overline{\Delta g_{(s, s+1)}} = \dfrac{\sum_{k \in \mathcal{Q}_{(s, s+1)}}\left( \Delta g^{k}_{(s, s+1)}\middle/\left(\sigma^k_{(s, s+1)}\right)^2\right)}{\sum_{k \in \mathcal{Q}_{(s, s+1)}} \left. 1\middle/\left(\sigma^k_{(s, s+1)}\right)^2\right.}\label{w_combine}
\end{equation}with its propagated error as
\begin{equation}
    \delta_{(s, s+1)}  = \sqrt{\left(\sum_{k\in\mathcal{Q}_{(s, s+1)}}\left(\sigma^k_{(s, s+1)}\right)^{-2}\right)^{-1}}\label{w_combine_err}
\end{equation}
where $\sigma^k_{(s, s+1)}$ is the standard deviation calculated from the time series of $\Delta g^{k}_{(s, s+1)}$ since the last update of the Wang-Landau incrementor in the EE simulation having sampling the $k$-th state set. From the averaged weight differences between all adjacent states, we can obtain a profile of alchemical weights that can be used to initialize the next iteration.

\subsection{Free energy calculations}
We term REXEE simulations composed of fixed-weight EE replicas as \textbf{fixed-weight REXEE} simulations, and those composed of weight-updating EE replicas as \textbf{weight-updating REXEE} simulations. For free energy calculation, the protocol used for the fixed-weight REXEE simulations is similar to those for HREX and fixed-weight EE simulations, albeit with extra consideration for overlapping states. For each state set, one should concatenate the trajectories from all replicas, truncate the non-equilibrium region~\cite{chodera2016simple} and then decorrelate the concatenated data. Then, for each replica in the fixed-weight REXEE simulation, one can use free energy estimators such as TI~\cite{kirkwood1935statistical}, BAR~\cite{bennett1976efficient}, and MBAR~\cite{shirts2008statistically} to calculate the alchemical free energies for different state sets. For the overlapping states, one can use Equations \ref{w_combine} and \ref{w_combine_err} to calculate the mean of the associated free energy differences $\overline{\Delta G_{(s, s+1)}}$ and the accompanying propagated error $\delta_{(s, s+1)}$, with $\Delta g^k_{(s, s+1)}$ replaced by $\Delta G^k_{(s, s+1)}$, the free energy difference computed by the chosen free energy estimator. In this context, $\sigma^k_{(s, s+1)}$ used in Equations \ref{w_combine} and \ref{w_combine_err} should be the uncertainty associated with  $\Delta G^k_{(s, s+1)}$ calculated by the estimator. Importantly, free energy differences involving overlapping states are likely to have smaller uncertainties because more uncorrelated samples can be collected given a larger pool of samples gathered from multiple replicas. 

Notably, in Equations \ref{w_combine} and \ref{w_combine_err}, we consider only the path of $0 \rightarrow 1 \rightarrow 2 \rightarrow \ldots \rightarrow s \rightarrow s+1 \rightarrow \ldots \rightarrow N$ for transitioning from the coupled state (state 0) to the decoupled state (state N) in the contexts of both weight combinations and free energy calculations. Other pathways, such as $0 \rightarrow 2 \rightarrow 4 \rightarrow \ldots \rightarrow s \rightarrow s+2 \rightarrow \ldots \rightarrow N$, as well as more complex or irregular paths, may also be viable paths for weight combinations and free energy calculations. It should be noted, however, that the resulting weight/free energy difference between states 0 and N, along with its associated uncertainty, will vary based on the chosen path. Although all paths connecting the end states yield valid results, we advocate for the approach that considers all adjacent states, as used in Equations \ref{w_combine} and \ref{w_combine_err}. The primary reason for this preference is that the path composed of only one-state moves offers the greatest flexibility, simplicity and can accommodate any degree of overlap between state sets. Moreover, in instances where the uncertainties $\sigma^k_{(s, s+1)}$ from different replicas $k$ are close, path selection will have minimal impact on the final values $\Delta g_{(1, N)}$ or $\Delta G_{(1, N)}$ and their respective uncertainties.

An alternative way to estimate the uncertainty of $\Delta G_{(s, s+1)}$ in Equation \ref{w_combine} is to perform bootstrapping, which is path-independent. In theory, it is also more accurate and rigorous than the propagated error in Equation \ref{w_combine_err}, but more computationally expensive. Specifically, in one bootstrap iteration, one can replicate the input dataset that is already concatenated, truncated, and decorrelated by drawing $N_\text{data}$ samples with replacement, where $N_\text{data}$ is the size of the dataset. From the sampled dataset, one can repeat the protocol above to get a free energy estimate. Then, the uncertainty can be estimated by taking the standard deviation of the free energy estimates obtained from a predetermined number of bootstrap iterations. 

\section{Methods} \label{methods}
\cprotect\subsection{The \verb|ensemble_md| Python package}
\verb|ensemble_md| is a pip-installable Python package that houses all algorithms required for implementing the REXEE approach. It is compatible with GROMACS starting from version 2022.5. As an adaptable wrapper around GROMACS functionalities, \verb|ensemble_md| provides a layer of high-level abstraction over the complexity of performing a REXEE simulation. Specifically, it offers several user-friendly CLIs, such as \verb|explore_REXEE|, \verb|run_REXEE| and \verb|analyze_REXEE|. The CLI \verb|explore_REXEE| solves Equation \ref{Nnrs} with additional constraints to efficiently enumerate all possible REXEE configurations, providing the user a quick overview of the parameter space for configuring the REXEE simulation. The CLI \verb|run_REXEE| streamlines the workflow of conducting a REXEE simulation by orchestrating the iterative cycle of GROMACS simulations. This includes preparing the simulation inputs for each iteration, launching simulations, and performing brief, on-the-fly data analysis for parameter adjustment between iterations. Lastly, the CLI \verb|analyze_REXEE| integrates an assortment of data analysis methods tailored for REXEE simulations. Tasks automated by  \verb|analyze_REXEE| include trajectory stitching, replica- and state-based transition analysis, time series analysis, Markov State Model (MSM)~\cite{husic2018markov, bowman2013introduction, scherer2015pyemma} analysis, data visualization, and free energy calculations. 

At its core, \verb|ensemble_md| implements basic functionalities that are central to the REXEE method. This includes methods for swapping input configurations between replicas, calculating the acceptance ratio, and managing the behavior of the next EE iteration by tweaking the input simulation parameters. It also provides different proposal weights and weight combination methods introduced in the section ``Theory''. By launching subprocess calls of the GROMACS executable, \verb|ensemble_md| eliminates the need to alter the source code of GROMACS, while ensuring negligible overhead in Python execution. Owing to the modularity of its core functionalities, \verb|ensemble_md| is also easily extensible. It provides building blocks for easy extension of the REXEE approach (e.g., by enabling customized swapping schemes), or formulating novel simulation approaches (e.g., by combining the replica exchange method with other enhanced sampling methods). In addition, the majority of \verb|ensemble_md|'s functions are agnostic of the MD simulation engine, paving the way for integration with other MD engines by merely adding a few engine-specific functions to handle different file extensions and data types. Lastly, \verb|ensemble_md| ensures quality through extensive unit testing, continuous integration, and comprehensive documentation.

\subsection{Simulations of anthracene}
To showcase the applicability of the REXEE approach, we applied the method to the calculation of the solvation free energy of anthracene, a selection made based on several compelling factors. Primarily, anthracene presents a balanced level of challenge for free energy methodologies---it is cheap enough to allow efficient exploration of diverse simulation parameter setups while still providing a sufficient challenge to rigorously test the efficacy of free energy methods. Furthermore, the absence of configurational metastable states in the anthracene system eliminates the possibility of having configurational free energy barriers orthogonal to the alchemical variable, simplifying the comparison of alchemical sampling among generalized ensemble methods, including EE, HREX, and the REXEE methods. Lastly, anthracene has been extensively studied in the work by Paliwal et al.~\cite{paliwal2011benchmark}, which offers a comparative reference. With this system, we compared the free energy estimates from fixed-weight REXEE simulations with different setups to the benchmark values obtained from HREX and fixed-weight EE simulations. We also compared REXEE and EE methods in their ability to converge alchemical weights in the weight-updating stage. All simulations described in the following subsections were performed with GROMACS 2022.5, with the aid of the package \verb|ensemble_md| for the REXEE simulations. To enable a straightforward comparison of computational costs, all simulations were carried out on identical computational architectures. Example simulation inputs, including initial configurations, topologies, and simulation parameters are available at \url{https://github.com/wehs7661/ensemble\_md/tree/master/ensemble\_md/data}.

\subsubsection{System preparation}
For the anthracene system, we selected one of the configurations from the study by Paliwal et al.~\cite{paliwal2011benchmark} as the initial configuration for the subsequent downstream investigation. We equilibrated the system in the NVT ensemble and then the NPT ensemble. Both equilibration processes were performed for 200 ps, with the velocity rescaling method~\cite{bussi2007canonical} used in both to maintain the reference temperature at 300K, and a Berendsen barostat~\cite{berendsen1984molecular} used in the latter to fix the pressure at 1 bar. Afterward, we carried out a 5 ns NPT MD simulation with a Parrinello-Rahman barostat~\cite{parrinello1980crystal, parrinello1981polymorphic} keeping the pressure at 1 bar. A switching function was used for the calculation of van der Waals interaction, where the switch started at 0.8 nm and the cutoff distance was set at 0.9 nm. For efficient calculations of long-range electrostatic interactions, the PME (particle mesh Ewald) method~\cite{essmann1995smooth} was used with a cutoff distance of 0.9 nm and a grid spacing of 0.1 nm. Bonds involving hydrogen bonds were constrained by the LINCS algorithm~\cite{hess1997lincs} with 2 iterative corrections. The highest order in the expansion of the constraint coupling matrix was specified as 12. Upon completion of the MD simulation, we extracted the configuration with the box volume closet to the average volume of the MD trajectory. This configuration then served as the input configuration for subsequent EE, HREX, and REXEE simulations elaborated in the following sections, as all these simulations were conducted in the NVT ensemble to avoid potential issues with $\lambda$ dependence of pressure. 

\subsubsection{Benchmark simulations}
To establish benchmark values of solvation free energy for comparison with the results from REXEE simulations, we performed EE and HREX simulations for the anthracene system, both in the NVT ensemble. Both benchmark simulations utilized 8 alchemical intermediate states in total, which is a convenient number to parallelize the replicas in the HREX simulation. These 8 alchemical states were used to only decouple the van der Waals interactions, as the anthracene model had zero charge on all atoms. Their coupling parameters were chosen to ensure sufficient overlap between adjacent states. To avoid singularities in standard Lennard-Jones potentials, soft-core interpolation between the end states was applied. The parameters $\alpha$ and $p$ in the soft-core potentials were specified as 0.5 and 1, respectively. 

For the EE benchmark, we adopted the prevalent two-stage protocol described in the Introduction section. In the weight-updating phase, we used the $1/t$ variant of the Wang-Landau algorithm~~\cite{belardinelli2007fast, belardinelli2007wang} to converge the alchemical weights for the 8 alchemical states, with the initial Wang-Landau incrementor set to 0.5 $k_{\text{B}} T$. Monte Carlo moves in the alchemical space were proposed every 100 integration steps with Metropolized-Gibbs sampling. We adopted the default value of 0.8 for the flatness ratio $R_{\text{flat}}$, a ratio between the histogram counts of a state and the histogram counts averaged over all states. A cutoff of 0.8 means the histogram is considered flat only if all states have an $R_{\text{flat}}$ value smaller than 0.8, and its reciprocal $1/R_{\text{flat}}$ larger than 0.8.  Whenever the histogram is considered flat, all histogram counts are reset to 0 with the Wang-Landau incrementor scaled by a scaling factor, which was set to 0.8 in our case. This weight-updating procedure ceased when the Wang-Landau incrementor fell below 0.001 $k_{\text{B}} T$, reaching what we term as the \textbf{Wang-Landau (WL) convergence}. The weights converged by the Wan-Landau algorithm were then used and fixed during the EE simulation in the production phase, which was performed for 200 ns. The fixed-weight EE simulation used exactly the same set of parameters as those used in the weight-updating stage, except that no weight-updating settings were specified. To determine the number of uncorrelated samples, we truncated the nonequilibrium regime and decorrelated the time series of the Hamiltonian difference between adjacent states~\cite{chodera2016simple} generated by the fixed-weight EE simulation. Then, we applied MBAR~\cite{shirts2008statistically} to the uncorrelated samples to compute the free energy difference between the coupled and decoupled states, which is the solvation free energy of anthracene. 

The HREX benchmark simulation used one replica for each of the 8 intermediate states. All 8 replicas were seeded with the same initial configuration and each of them was performed for 25 ns, summing up to the same total simulation time of 200 ns as the production phase of the EE benchmark simulation. The parameters used in the HREX benchmark simulation are identical to those in the fixed-weight EE simulation, except that no alchemical weights were assigned and no expanded ensemble settings were used. For free energy calculations, we truncated and decorrelated the time series of Hamiltonian differences for each replica individually, then concatenated data from different replicas for MBAR~\cite{shirts2008statistically} calculations. 

\subsubsection{Fixed-weight REXEE simulations}
To make the comparison between REXEE simulations and the benchmark simulations straightforward, we used the same 8 alchemical intermediate states ($N=8$) for all REXEE simulations. All EE replicas of all REXEE simulations have the same set of parameters as the EE benchmark simulation, except that the different EE replicas in REXEE were restricted to different state sets.

\begin{figure}[ht]
    \centering
    \includegraphics[width=\textwidth]{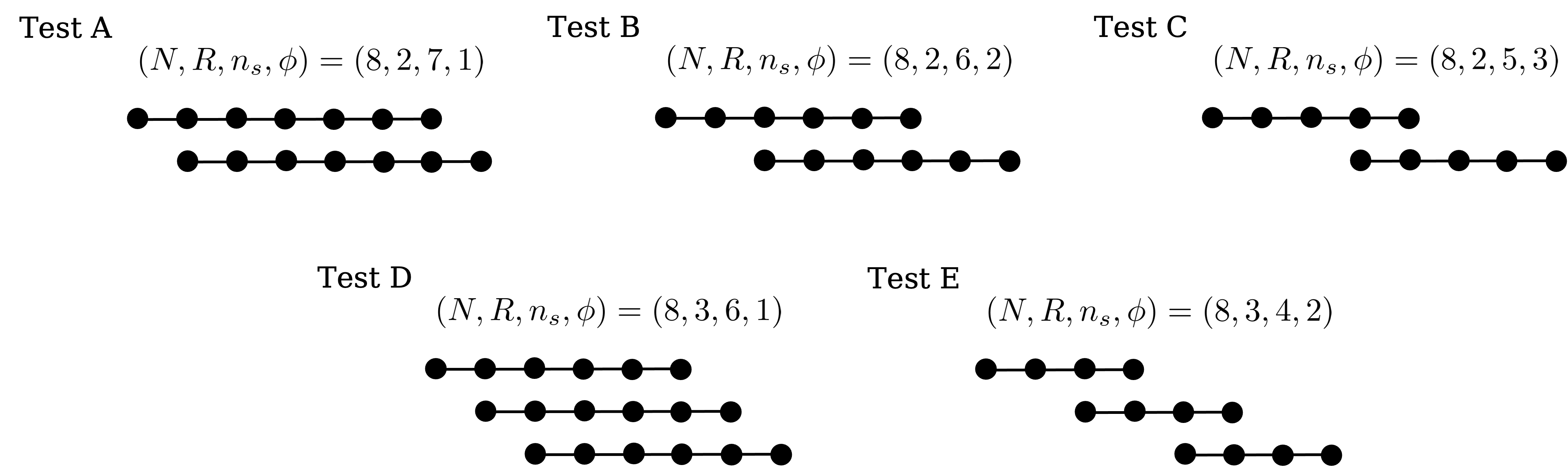}   
    \caption{The schematic representations of REXEE configurations of Tests A to E for anthracene solvation free energy calculation. The black dots represent alchemical states and different rows of dots represent different replicas.}
    \label{REXEE_config}
\end{figure}

With the anthracene system, we tested different setups of fixed-weight REXEE simulations and classified them into two groups. Group 1 includes Tests 1 to 3, which test the effect of different frequencies for exchanging replicas, with the corresponding simulation lengths per iteration (i.e., exchange period) of 4 ps, 10 ps, and 100 ps, respectively. All three tests were configured with $(N, R, n_s, \phi)=(8, 4, 5, 1)$. In parallel, Group 2 includes Tests A to E, which explores 5 different $(N, R, n_s, \phi)$ combinations varying the number of replicas and the level of overlap. All 5 tests were performed at a fixed exchange frequency of swapping replicas every 4 ps, a frequency shown to lead to faster mixing in the sampling space and decent free energy estimates in Group 1 (see the section ``Results and Discussion''). The 5 considered $(N, R, n_s, \phi)$ combinations include $(8, 2, 7, 1)$, $(8, 2, 6, 2)$, $(8, 2, 5, 3)$, $(8, 3, 6, 1)$, and $(8, 3, 4, 2)$. (See Figure \ref{REXEE_config}.) All REXEE simulations in Groups 1 and 2 have the same effective simulation lengths as those of benchmarks, which are 200 ns. All simulations adopted the exhaustive exchange proposal scheme and were all initiated with the same set of alchemical weights obtained in the weight-updating phase of the EE benchmark simulation. 

To assess the performance of the REXEE method, we compared the free energy estimates of all 8 REXEE simulations with the benchmark simulations. Specifically, for each state set, we stitched all Hamiltonian time series, then truncated and decorrelated the combined time series~\cite{chodera2016simple} for MBAR~\cite{shirts2008statistically} calculations. The free energy profiles obtained from different state sets are then combined as described in the Theory section. Furthermore, from each demuxed trajectory, we quantified the sampling speed of REXEE simulations in both the replica space and the state space. If Hamiltonian replica exchange is being used as a configurational sampling technique, then faster sampling in state space is a necessary (though not sufficient) requirement for improved configurational sampling. Specifically, we calculated the replica-space relaxation time ($\tau_r$) given by below:\begin{equation}
    \tau_r = \frac{\tau}{1-\lambda_2}
\end{equation} where $\tau$ is the exchange period and $\lambda_2$ is the second-largest eigenvalue calculated from the replica-space transition matrix. As a measure of the sampling speed in the replica space, $\tau_r$ provides an estimate of the simulation time required for the autocorrelation function of the replica index to decay to $1/e$ of its initial value, and a shorter replica-space relaxation time is indicative of faster mixing in the replica space. As for the state-space sampling speed, we adopted two straightforward metrics, including the correlation time of the state index ($\tau_{\lambda}$) and the number of round trips ($n_r$) in the alchemical space. Notably, for the state index correlation time $\tau_{\lambda}$, we report averages over $R$ available trajectories with uncertainties being the standard deviation. For the number of round trips $n_r$, we report the sum over all $R$ trajectories, so that all values are based on the same simulation length. Its uncertainty was estimated as $1/\sqrt{R}$ times the standard deviation calculated from the $R$ trajectories, as the uncertainty should scale with $1/\sqrt{R}$ as the sample size scales with $R$. Lastly, we estimated the uncertainty of the replica-space relaxation time $\tau_{\lambda}$ by applying bootstrapping to synthesized replica-space transition matrices that mock the empirical transition matrix. For metrics applicable to the benchmark simulations (i.e., $\tau_{\lambda}$ and $n_r$), we also applied them to assess the sampling speed of the benchmark simulations. Notably, the comparison of sampling speed and free energy calculations were not only carried out between the REXEE simulations and the benchmark simulations, but also between the REXEE simulations with different setups. 

\subsubsection{Weight-updating REXEE simulations}
In addition to fixed-weight REXEE simulations, we also performed 6 weight-updating REXEE simulations, which are the 6 combinations of three exchange periods (4 ps, 10 ps and 100 ps) and whether the weight combination scheme based on simple averages was used or not. All 6 simulations were configured by $(R, n_s, \phi) = (4, 5, 1)$, a setup that has a reasonable overlap between replicas and resulted in a decent performance of the simulation given a sufficiently high exchange frequency (see the section ``Results and Discussion''). All 6 simulations were scheduled for 50 ns each (i.e., 200 ns in total) and they all used the exhaustive exchange proposal scheme for swapping the coordinates between replicas. 

To assess REXEE's ability to converge alchemical weights, we adopted two metrics. First, we measured the WL convergence time for each test, which refers to the time it takes for all replicas to converge the weights for their respective state set according to the Wang-Landau algorithm's criteria. Second, for each REXEE simulation, we calculated the root-mean-squared error (RMSE) between the free energy profile averaged over the period since the last update of the Wang-Landau incrementor and the reference profile calculated from the EE benchmark simulation. Due to the dynamic nature of the weight-updating phase, we performed all 6 simulations in 3 replicates and report the average to lower the influence of noise, with the uncertainty being the standard deviation. The metrics of each test were then compared with those averaged over 3 replicates of weight-updating EE simulation that used the $1/t$ variant of the Wang-Landau algorithm to converge the alchemical weights for all states.

\subsection{Simulations of CB7-10 host-guest binding complex}
To further understand the REXEE approach, we applied the method to the binding free energy calculation of the host-guest binding complex CB7-10. CB7-10, composed of a cucurbit[7]uril (CB7) as the host and a guest ligand, is one of the binding complexes in SAMPL4 SAMPLing challenge.~\cite{muddana2014sampl4}  In contrast to anthracene, CB7-10 has at least three predominant metastable states, including the unbound state and two symmetric binding sites on each side of the host.  This presents a more complex scenario than solvation free energy calculations, but not to the extent that comprehensive sampling in configurational space becomes overly challenging, given that its slow degrees of freedom are not strictly orthogonal to the alchemical direction. Furthermore, CB7-10 offers the advantage of testing the accuracy of not only solute-water but also solute-solute interactions, making it an ideal candidate for evaluating free energy methods in practical contexts. With this system, we assessed REXEE simulations with various setups, evaluating the sampling speed in both the alchemical space and the configurational space, as well as the accuracy of free energy calculations and the convergence of alchemical weights in the weight-updating stage. All REXEE simulations were compared with each other and also with the EE benchmark simulation. Free energy estimates from these simulations were compared with the values reported in the study by Monroe et al.~\cite{monroe2014converging}, which also employed the EE method. All simulations were carried out in the NVT ensemble using GROMACS 2022.5 on identical computational architecture, with the package \verb|ensemble_md| employed specifically for REXEE simulations. Example simulation inputs, including initial configurations, topologies, and simulation parameters are available at \url{https://github.com/wehs7661/ensemble\_md/tree/master/ensemble\_md/data}.

\subsubsection{System preparation}
Starting from the coordinate file provided in SAMPL4 SAMPLing challenge repository~\cite{IorgaMobley2021} and the topology file adopted in the study by Monroe et al.~\cite{monroe2014converging}, we solvated the system in a cubic box with 1.5 nm between the solute and box edges. After charge neutralization with three chloride ions, the system was then proceeded with energy minimization followed by NVT equilibration, NPT equilibration, and finally, a 5 ns NPT MD simulation. The structure whose box volume was closest to the average volume of the MD trajectory was extracted to serve as the input for REXEE and EE benchmark simulations, which were all performed in an NVT ensemble. All parameters used in these steps are the same as those used for the preparation of the anthracene system, except that switching functions were used in this case for both the calculations of van der Waals and electrostatic interactions in the last step, with the range for switching being 0.85 to 0.9 nm and 0.89 to 0.9 nm, respectively. 

\subsubsection{Decomposition of the binding free energy calculation}
\begin{figure}[ht]
    \centering
    \includegraphics[width=\textwidth]{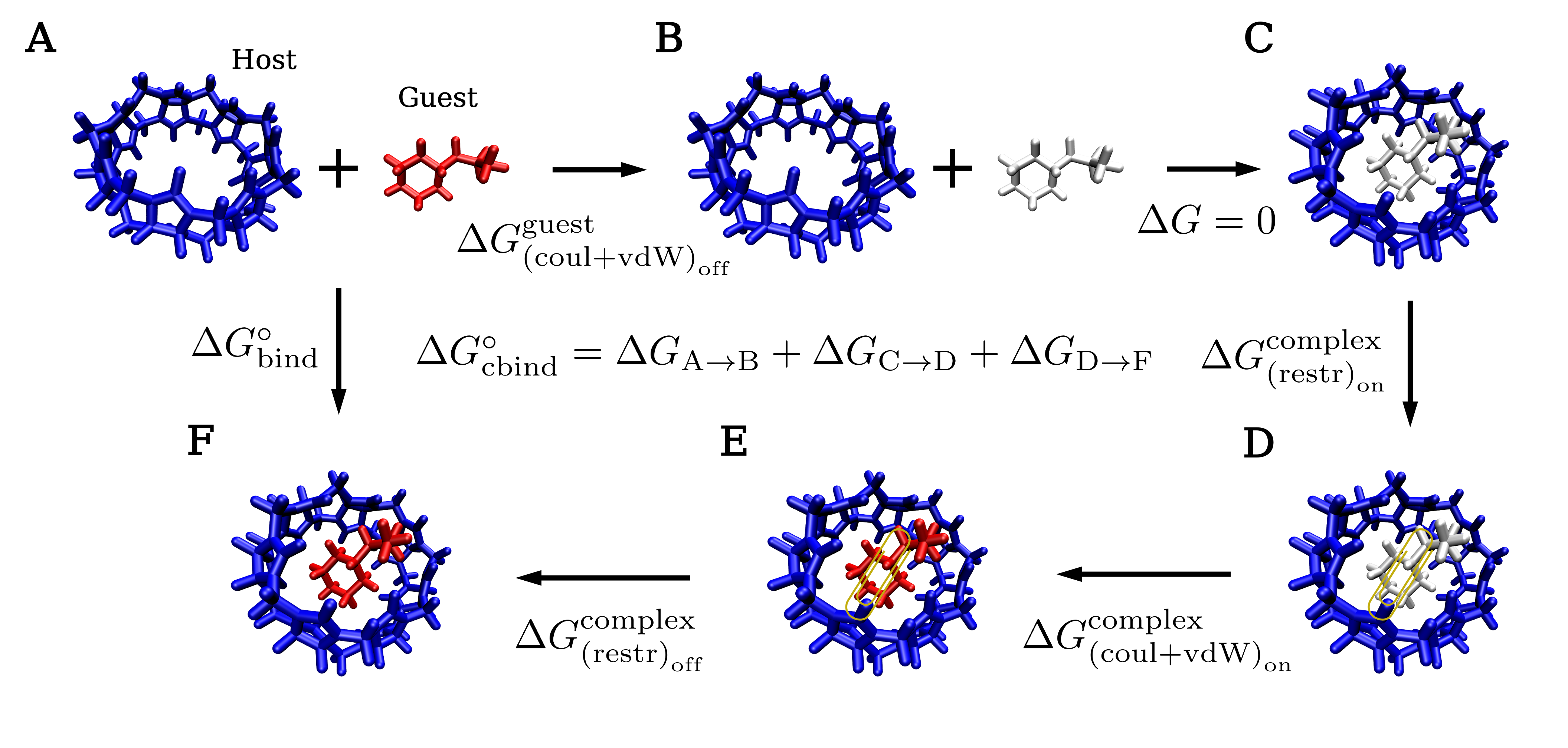}   
    \caption{The double decoupling thermodynamic cycle for the binding free energy calculation of the CB7-10 host-guest binding complex. The host molecule is colored in blue, while the guest ligand is colored in either white (fully decoupled from the host) or red (fully coupled with the host). The paper clips (yellow) in states D and E represent the distance restraint applied between the host and guest molecules.}
    \label{thermo_cycle}
\end{figure}

We used the double decoupling method~\cite{gilson1997statistical} with the thermodynamic cycle shown in Figure \ref{thermo_cycle} to decompose the binding free energy calculation of CB7-10 binding complex. Starting from state A, we split the entire cycle into two alchemical processes (from states A to B and from states D to F) bridged by the transfer of the ligand into the binding cavity of the host molecule (from states B to C) and the application of a distance restraint (from states C to D). The associated free energy difference of each alchemical process was calculated from a single fixed-weight EE or REXEE simulation. The simulation for the transition from states A to B, which we term the ``solvent simulation'', gradually decouples the ligand from its surroundings. After the ligand is fully decoupled, we can freely move it into the binding cavity of the host molecule without any energy penalty, so the associated free energy difference is 0. Then, it is of common practice to apply a distance restraint between the host and guest molecules to prevent the guest ligand from drifting away, which effectively shortens the decorrelation time. The change in the free energy associated with this process can be calculated analytically:
\begin{equation}
    \Delta G^{\text{complex}}_{\text{(restr)}_{\text{on}}}= -k_B T \ln \left [ \frac{1}{V_0}\left(\left(\frac{2 \pi k_B T}{K}\right)^{3/2} + \frac{8\pi r_0 k_B T}{K} + 2\pi r_0^2 \left(\frac{2\pi k_B T}{K}\right)^{1/2}\right)\right]\label{correction}
\end{equation}
where $r_0$ is the reference distance, $K$ is the force constant and $V_0$ is the molecular volume ($1.6605$ $\mathrm{nm^3}$) corresponding to the $1$ $\mathrm{mol/L}$ reference concentration. The thermodynamic cycle can then be closed by the so called ``complex simulation'' going from states D to F, which turns back on the non-bonded interactions and switches off the distance restraint. Finally, the binding free energy $\Delta G^{\circ}_{\text{bind}}$ can be calculated as the sum of the free energy differences associated with the process going from states A through B, C, D, E to F. To simplify the comparison between the EE and REXEE methods, we only compared the methods in the complex simulation for the calculation of $\Delta G_{\text{D}\rightarrow\text{F}}$, which is typically the more challenging part of binding free energy calculations. We therefore only used the EE method to calculate $\Delta G_{\text{A}\rightarrow\text{B}}$ from the solvent simulation for all comparisons.
\subsubsection{Solvent simulation}
As described above, the solvent simulation, which samples the transition from states A to B in Figure \ref{thermo_cycle}, was performed using only the EE method with the common two-stage protocol. We adopted the simulation parameters used in the study by Monroe et al. wherever possible for a more straightforward comparison of the free energy estimates. This includes the specification of the alchemical path and the settings for Monte Carlo moves in the state space, alongside the parameters for both the Wang-Landau algorithm and the soft-core potential. Specifically, 40 alchemical states in total were used, 21 of which were used to decouple the Coulombic interactions and the others for the van der Waals interactions. In the weight-updating stage, the $1/t$ variant of the Wang-Landau algorithm was used to adaptively estimate the weight for each alchemical state, with an initial Wang-Landau incrementor of 10 $k_{\text{B}}T$, a flatness ratio $R_{\text{flat}}$ of 0.8, a scaling factor of 0.7, and a cutoff of 0.001 $k_\text{B}T$ for halting the weight-updating process. Monte Carlo moves in the state space were attempted every 100 integration steps using the Metropolized-Gibbs sampling method. The parameters $\alpha$ and $p$ for the soft-core potential were set to 0.5 and 1, respectively. The converged weights were used to seed the 200 ns production run, from which the data was extracted and fed to the MBAR estimator for free energy calculations after proper truncation and decorrelation. Both simulations were conducted in the NVT ensemble to avoid any potential issues with $\lambda$ dependence of pressure. This is as opposed to the EE simulations performed in Monroe et al.'s study, which were performed in the NPT ensemble, as changes in GROMACS since that time made it impossible to run the simulations in the NPT ensemble.  However, in an aqueous solution at 1 atm, the difference between Helmholtz and Gibbs free energies is within statistical error if run at properly corresponding pressure/volumes.

\subsubsection{Benchmark simulations}
We performed EE simulations with the two-stage protocol to calculate a benchmark value for the free energy difference $\Delta G_{\text{D}\rightarrow\text{F}}$, to which the free energy estimates from the REXEE simulations were compared. In the weight-updating stage, we set the initial Wang-Landau incrementor to 10 $k_{\text{B}}T$ to allow more rapid equilibration given the large free energy difference in desolvation of a charged complex. A $\lambda$-dependent distance restraint was applied between the ligand and the host molecules to reduce the movement of the ligand in the unbound state, which could otherwise explore the entire simulation box. The force constant was set to 1000 $\mathrm{kJ/mol/nm^2}$ and the initial distance between the centers of mass of the two entities, $r_0=0.428$ $\mathrm{nm}$, was used as the reference distance, at which structures correspond to the bound state. All the other simulation parameters used in the weight-updating EE simulation are the same as those used in the solvent simulation in the weight-updating stage. The fixed-weight EE simulation for production used the same set of parameters as the weight-updating simulation except that no weight-updating parameters were specified and the weights were initialized at the converged values obtained in the weight-updating stage. Similarly, after truncation and decorrelation, we applied MBAR to compute the free energy difference $\Delta G_{\text{D}\rightarrow\text{F}}$.

To assess the binding and unbinding dynamics of the host-guest binding complex, we performed a clustering analysis for the fixed-weight EE simulation and used it as a benchmark for later comparison with REXEE simulations. We first removed periodicity and jumps of molecules across the simulation box from the trajectory, centered the binding complex, and performed clustering using the single linkage algorithm for the fully coupled configurations, with an RMSD cutoff of 0.13 nm, a value also adopted in the study by Monroe et al.~\cite{monroe2014converging} Given the trajectory with each frame assigned to a cluster, we then calculated the number of flips between the two most dominant clusters, which were expected to be the guest molecule bound to the two different portals of the host molecule, to quantify the sampling efficiency in the configurational space. 

\subsubsection{Fixed-weight REXEE simulations}
We performed 8 fixed-weight REXEE simulations for the CB7-10 host-guest binding complex to sample the same 40 alchemical states as those defined in the EE benchmark simulation. These simulations, denoted as Tests 1 to 8, explore 8 different REXEE configurations, with the $(N, R, n_s, \phi)$ combination including $(40, 4, 13, 9)$, $(40, 4, 19, 7)$, $(40, 4, 37, 1)$, $(40, 6, 10, 6)$, $(40, 6, 15, 5)$, $(40, 6, 35, 1)$, $(40, 8, 12, 4)$, and $(40, 8, 33, 1)$. These configurations were chosen such that for each $R$ value (number of replicas) of 4, 6, 8, the configurations with the highest and lowest overlap between adjacent state sets were included, with additions of an intermediate overlap for $R=4$ and 6 to capture a broader insight into the performance spectrum of these setups. All 8 tests use the exhaustive exchange proposal scheme to propose exchanges every 4 ps, a frequency that is highest among those tested in the anthracene simulations and leads to the highest sampling efficiency in the alchemical space (see the section ``Results and Discussion''). All tests have the same aggregate simulation length of 200 ns and used the same set of parameters as utilized in the EE benchmark simulation. They were all initiated with the same weights obtained from the weight-updating phase of the EE benchmark simulation. We utilized the same data analysis protocol used for the anthracene fixed-weight simulations to analyze the CB7-10 simulations, with an exception in the time series decorrelation protocol for free energy calculations. Specifically, we found that the data decorrelation method~\cite{chodera2016simple} occasionally underestimated the statistical inefficiency of CB7-10 trajectories, and we therefore applied the geometric mean of the statistical inefficiency over all trajectories to average out the uncertainty intrinsic to the method.

To examine the sampling efficiency in the alchemical space, we adopted the same protocol as the one used for analyzing fixed-weight REXEE simulations of anthracene. To assess the configurational sampling of the CB7-10 REXEE simulations, we first stitched and recovered a continuous trajectory for each starting configuration, and applied the same protocol for clustering analysis described in the previous section to get the flipping rate for each simulation. Given that the aggregate length of each REXEE simulation is the same as the length of the benchmark EE simulation, for each REXEE simulation, we summed up the number of flips across all trajectories and compared the sum with the flip count calculated from the EE benchmark simulation.

\subsubsection{Weight-updating REXEE simulations}

In contrast to the weight-updating REXEE simulations performed for the anthracene system, which focused on one single REXEE configuration and investigated different exchange frequencies and the use of weight combinations, the 8 weight-updating REXEE simulations performed for the CB7-10 system explored different REXEE configurations using the same exchange frequency. Referred to as Tests 1 to 8, these tests examine the same 8 REXEE configurations as those explored in the 8 tests of fixed-weight REXEE simulations described in the previous section. They did not use any weight combination given that the results from the anthracene system showed that weight combinations generally impeded the convergence of weights. Additionally, given that the exchange frequency did not have a noticeable influence on the weight convergence in the anthracene system, we adopted an intermediate swapping period of 10 ps in the exhaustive exchange proposal scheme. (See the section ``Results and Discussion''.) Then, we again utilized the WL convergence time and RMSE with respect to the EE benchmark simulation to assess the weight convergence of REXEE simulations. All REXEE simulations were done in 3 replicates and averages across replicates are reported with standard deviations as uncertainties. For comparison, we also performed 3 replicates of weight-updating EE simulations and assessed them with the same metrics. 

\section{Results and Discussion}
\subsection{Simulations of anthracene}
\subsubsection{Benchmark simulations}
With the weights fixed at the values converged by the Wang-Landau algorithm, the solvation free energy of anthracene estimated by the benchmark EE simulation was 3.502 $\pm$ 0.178 $\mathrm{k_BT}$, which is statistically consistent with the estimate of 3.411 $\pm$ 0.067 $k_{\text{B}}T$ derived from the HREX benchmark simulation with the same effective simulation length. While these two values do not agree within uncertainty to the value reported in the work by Paliwal et al., ~\cite{paliwal2011benchmark} this disagreement does not affect our demonstration of the REXEE method, since internal consistency is achieved between the EE and HREX benchmarks and between results from the REXEE simulations and the benchmarks, as we show in the next subsection. We have included a discussion about the potential reasons for this discrepancy in free energy estimates in Section 3.1 in the Supporting Information.

Interestingly, the EE benchmark simulation showed shorter state index correlation time and more round trips than the HREX benchmark simulation (see Figure \ref{sampling_speed}). This can be attributed to the fact that EE simulations have provably higher exchange acceptance rates than HREX simulations given the same alchemical path~\cite{park2008comparison}.  

\subsubsection{Fixed-weight REXEE simulations}
\begin{figure}[ht!]
    \centering
    \includegraphics[width=\textwidth]{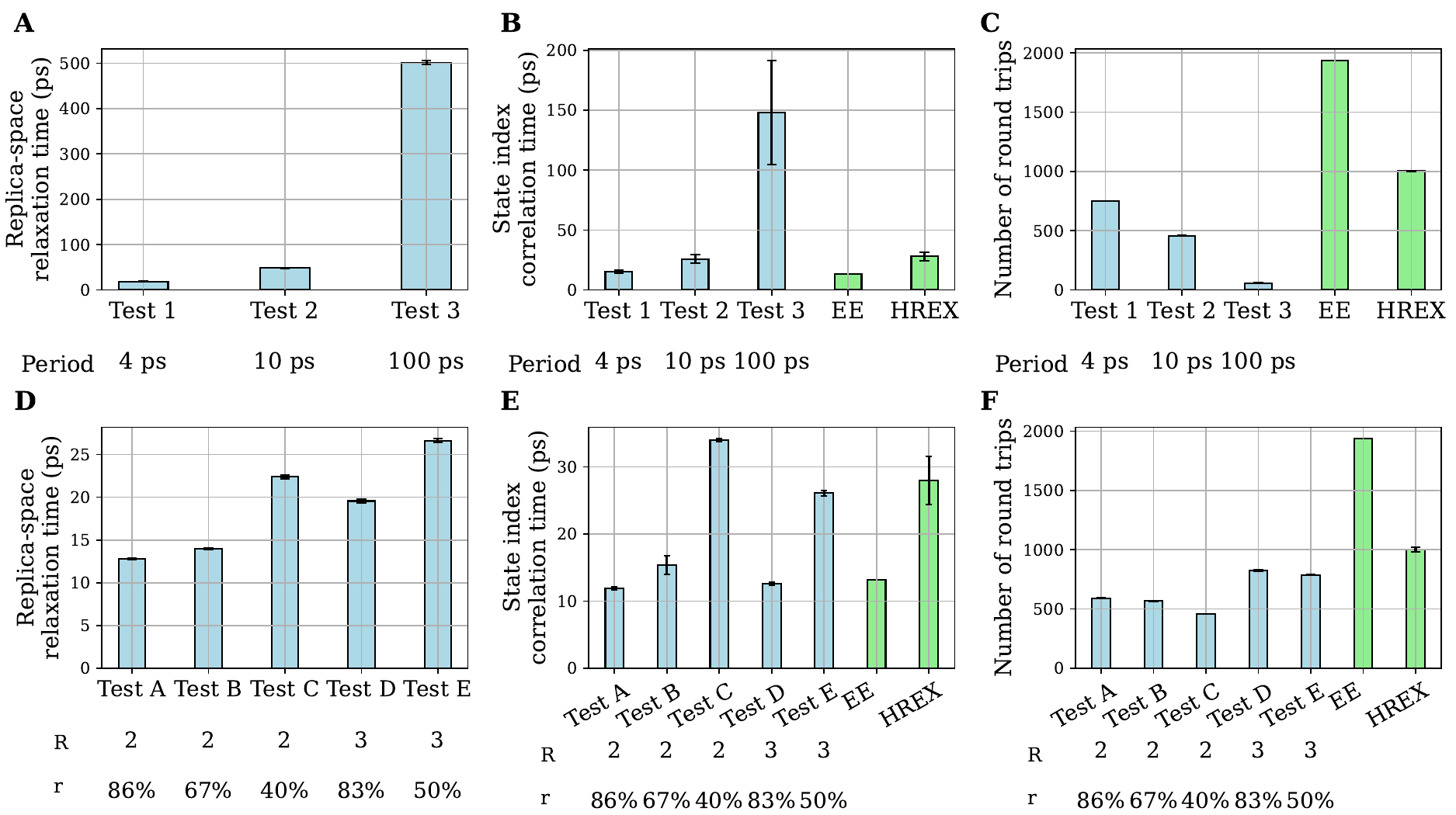}
    \caption{Different metrics for assessing the sampling speed for all fixed-weight REXEE simulations of anthracene, with parameters differing between tests annotated below the x-axis for easier comparisons. (In panels D to F, $R$ and $r$ represent the number of replicas and the overlap ratio, respectively.) Results from the benchmark simulations are in light green. The uncertainties of each metrics were calculated using the protocol described in section ``Methods'', with some error bars being too small to be visible. Overall, the figure shows that faster sampling in the replica space and the state space can be achieved by faster exchanges (shorter exchange periods) or higher state overlaps between adjacent state sets. (A) Replica-space relaxation time from tests in Group 1. (B) State index correlation time from tests in Group 1. (C) Total number of round trips from tests in Group 1. (D) Replica-space relaxation time from tests in Group 2. (E) State index correlation time from tests in Group 2. (F) Total number of round trips from tests in Group 2.}
    \label{sampling_speed}
\end{figure}

\begin{figure}[ht]
    \centering
    \includegraphics[width=\textwidth]{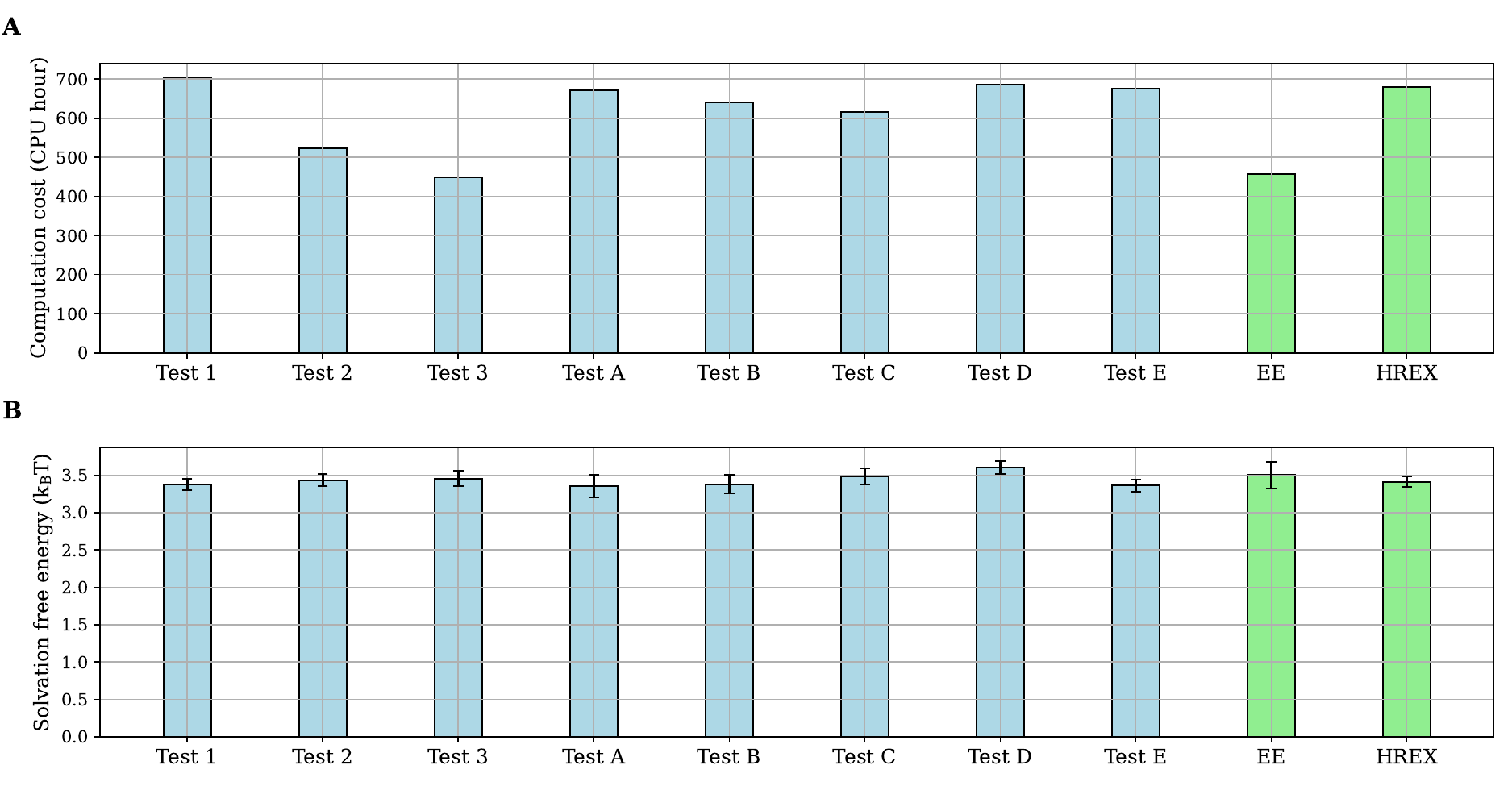}   
    \caption{(A) The computation costs (in CPU hours) of fixed-weight REXEE, fixed-weight EE, and HREX simulations. Tests 1 to 3 from Group 1, which respectively adopted an exchange period of 4 $\mathrm{ps}$, 10 $\mathrm{ps}$, and 100 $\mathrm{ps}$, exhibit decreasing computational costs due to less frequent simulation initialization. Additionally, Tests A to E in Group 2, which adopt the same exchange period (4 $\mathrm{ps}$) but different REXEE configurations, incurred similar computational costs, with some stochastic variability. (B) The estimates of the solvation free energy calculated from fixed-weight REXEE, fixed-weight EE, and HREX simulations. All REXEE tests, regardless of the exchange frequency and REXEE configuration, produced free energy estimates consistent with both EE and HREX benchmarks. The error bars of the free energy estimates were statistical errors calculated by the MBAR estimator.}
    \label{cost_free_energy}
\end{figure}

As hypothesized, faster sampling in either replica or state space can be achieved by both higher exchange frequency and larger overlap ratio, as assessed by metrics including the replica-space relaxation time, state index correlation time, and the total number of round trips. Specifically, panels A, B, and C in Figure \ref{sampling_speed} compare these three metrics between REXEE simulations with different exchange frequencies. We find that higher exchange frequency correlates with faster replica-space and state-space sampling, as shown by a shorter replica-space relaxation time, shorter state index correlation time, and more round trips in the state space. Notably, the observation of improved sampling efficiency in the state space with faster exchanges is consistent with earlier studies~\cite{sindhikara2008exchange, sindhikara2010exchange} that demonstrated enhanced mixing in pure replica exchange simulations with higher exchange frequencies. Interestingly, Test 1, distinguished by the highest exchange frequency, exhibited a shorter state index correlation time than the HREX benchmark simulation and was on par with the EE benchmark. Using the same high exchange frequency, tests in Group 2 also generally show a shorter state index correlation time than the HREX benchmark (see the bars corresponding to Tests A, B, and D in Figure \ref{sampling_speed}E.) This shows that REXEE can preserve the advantage of faster mixing as seen in the EE benchmark, which is unsurprising given that the state-space sampling in a REXEE simulation is conducted by the EE replicas. However, the use of these higher exchange frequencies demanded more computational resources, as can be seen from the comparison between Tests 1 to 3 in Group 1 in Figure \ref{cost_free_energy}A. This is a natural outcome of requiring more iterations to reach the simulation length equivalent to other tests, which necessitate frequent simulation initialization and result in a longer total GROMACS start time. Still, these additional overheads are within a reasonable range as they did not make REXEE simulations more expensive than the HREX benchmark simulation. 

As mentioned, tests in Group 2 propose an exchange every 4 ps, a frequency shown in Group 1 that allowed for rapid alchemical sampling without introducing an exorbitant computational cost relative to the HREX benchmark simulation. As a result, panel D in Figure \ref{sampling_speed} reveals a clear trend: given the same number of replicas, increasing the overlap between adjacent state sets accelerates mixing in the replica space, as readily observable in comparisons between Tests A, B, and C and between Tests D and E in panel D. This enhanced replica-space mixing, which we attribute to the more swappable pairs at exchanges given higher state overlap, in turn contributes to the acceleration of the state-space sampling as well (see Figure \ref{sampling_speed}E). Notably, the tests in Group 2, though varied in their REXEE configurations, shared a similar computational cost given the same exchange frequencies (see Figure \ref{cost_free_energy}A). While there is a noticeable trend that tests with higher overlap incur higher computational costs, the difference is generally marginal. 

We infer that the boundary of each state set can create an intrinsic barrier for the system to reach states outside the current state set it is sampling, as observed that all REXEE simulations exhibit fewer round trips than the EE and HREX benchmark simulations. While a higher exchange frequency helps diminish the barrier by increasing the flux across the boundaries (see Figure \ref{sampling_speed}C for Tests 1 to 3 in Group 1), different extents of overlap have a limited effect (see Figure \ref{sampling_speed}F). The importance of increasing the number of round trips usually lies in the pursuit of enhanced configurational sampling, especially when the fully coupled and decoupled states favor different metastable states in the configurational space, such as the bound and unbound state of a binding complex. Since the anthracene system does not have long-lasting configurational metastable states, the number of round trips does not directly influence the accuracy of free energy calculations. In fact, Figure \ref{cost_free_energy}B confirms the robustness of the REXEE method, as all REXEE simulations provided estimates of the solvation free energy of anthracene statistically consistent with both the EE and HREX benchmarks. Interestingly, the trend exhibited by Tests 1, 2, and 3 suggests that a faster exchange frequency results in lower uncertainty in free energy calculations. However, from Tests A to E there is no discernible trend of how the REXEE configuration (i.e., amount of overlap or number of replicas with fixed total simulation time) influences the accuracy of the solvation free energy estimate. 


\subsubsection{Weight-updating REXEE simulations}
\begin{figure}[ht!]
    \centering
    \includegraphics[width=\textwidth]{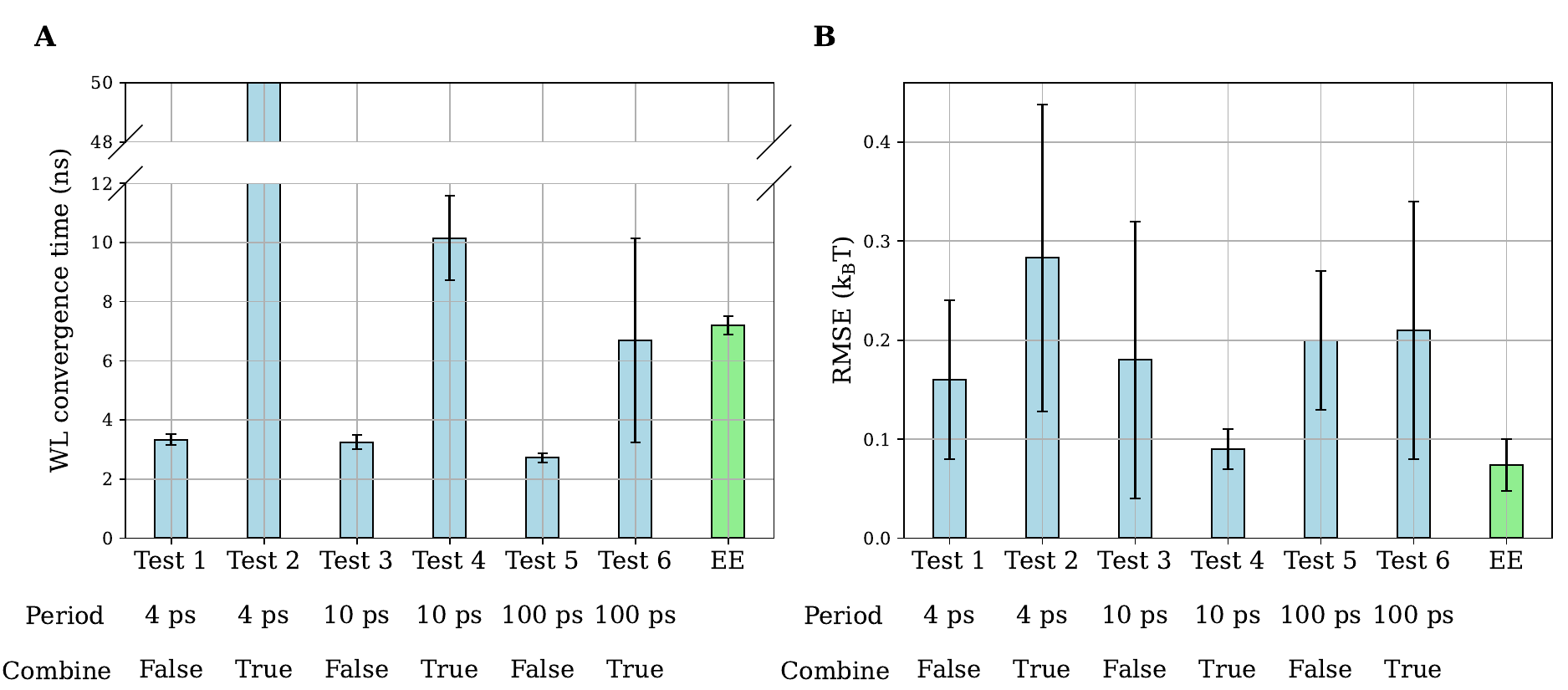}   
    \caption{(A) The Wang-Landau (WL) convergence time for all weight-updating REXEE simulations and the weight-updating EE simulation. For Test 2, 50 ns is reported because the simulation did not converge weights for all replicas within the scheduled length of 50 ns. (B) The RMSE relative to the reference free energy profile for all weight-updating REXEE simulations and the weight-updating EE simulation. For both panels, parameters differing between tests are annotated below the x-axis for easier comparisons, including the exchange period and whether the weight combination was used. The reported values are averages over 3 replicas, with the uncertainty being the standard deviation. The figure shows that using weight combination schemes did not shorten the WL convergence time nor effectively lower the RMSE value. Additionally, it shows that the exchange frequency has a limited effect on both the WL convergence time and RMSE value.}
    \label{t_equil}
\end{figure}
To assess the ability of the REXEE method in weight convergence, we compare different weight-updating REXEE simulations with weight-updating EE simulations in terms of the WL convergence time and the RMSE relative to the reference free energy profile. As a result, Figure \ref{t_equil}A shows that simulations that did not apply weight combination have a WL convergence time of 2 to 3 ns, while their counterparts that combine weights across replicas significantly lengthened the weight-converging process, with Test 2 notably failing to reach the criteria of the Wang-Landau algorithm for weight convergence within the scheduled simulation length of 50 ns. Taking this with the observation from Figure \ref{t_equil}B that weight combination did not reduce the RMSE values, we conclude that weight combinations based on simple averages did not introduce any advantage in weight convergence. We additionally tested more complicated weight combination schemes with different exchange frequencies, including weight combinations based on inverse-variance weighted means (using Equations \ref{w_combine} and \ref{w_combine_err}) and weight combinations with histogram corrections or weight corrections, but none of them outperformed simulations that did not use any weight combination scheme in terms of both WL convergence time and RMSE values. 

We reason that this surprising result that combining weights does not improve convergence is because the exchanges of coordinates between replicas have already caused each replica to visit all of the configurations that started with different replicas, and thus have ``seen'' the different configurations and incorporated them into the accumulated weights. Therefore, additionally combining weights across replicas may not provide any additional advantage. This is distinct from multiple replica metadyanmics, where each ``walker'' only samples a single configurational ensemble, and thus can benefit from weight information from alternate configurations. In addition, small changes in weights can drastically affect sampling, as state probabilities are exponential in the free energy differences between states. If one of the weights being combined is particularly bad, it will disrupt sampling for the other weights, and will therefore lower the convergence rate. This may also explain the trend in WL convergence time revealed in the comparison between Tests 2, 4, and 6 in Figure \ref{t_equil}A, where a higher exchange frequency led to more frequent ``contamination'' with any poorly converged weights and thus harmed the overall weight convergence. This is in contrast to the negligible impact of the exchange frequency on the WL convergence time as revealed by the comparison between Tests 1, 3, and 5, where weight combination was not used and different exchange frequencies were employed. Notably, although these three REXEE simulations achieved weight convergence in less wall time compared to weight-updating EE simulations, they incurred slightly higher computational costs due to running four replicas, as opposed to the single replica in the EE simulations. Lastly, Figure \ref{t_equil}B shows that the weights of all tests converged to values statistically consistent with those obtained from weight-updating EE simulations. Overall, for anthracene, it appears that the performance of weight convergence is relatively independent of REXEE parameters as long as no weight combination scheme is applied. 

\subsection{Simulations of CB7-10 host-guest binding complex}
\subsubsection{Benchmark simulations}
The benchmark EE simulation, which is 200 ns in length and had alchemical weights fixed at the values obtained from a weight-updating simulation, estimated the free energy difference $\Delta G_{\text{D}\rightarrow\text{F}} = \Delta G^{\text{complex}}_{\text{(restr)}_{\text{off}}} + \Delta G^{\text{complex}}_{\text{(coul + vdW)}_{\text{off}}}$ as $-645.773 \pm 0.050$ $\mathrm{k_BT}$. Additionally, the free energy difference $\Delta G_{\text{A}\rightarrow\text{B}} = \Delta G^{\text{guest}}_{\text{(coul + vdW)}_{\text{off}}}$ was estimated as $622.180 \pm 0.050$ $\mathrm{k_BT}$ by the solvent simulation. With the correction term $\Delta G_{\text{C}\rightarrow\text{D}}=\Delta G^{\text{complex}}_{\text{(restr)}_{\text{on}}}$ calculated as $2.260$ $\mathrm{k_BT}$ using Equation \ref{correction}, this leads to an estimate of the binding free energy $\Delta G^{\circ}_{\text{bind}}$ as $-21.33 \pm 0.07$ $\mathrm{k_BT}$. This value is statistically inconsistent consistent with the binding free estimate reported in the work by Monroe et al.~\cite{monroe2014converging}, mainly due to differences between GROMACS versions and that an incorrect correction term $\Delta G_{\text{C}\rightarrow{D}}$ was used in the reference work. A more detailed discussion about the disagreement between our estimate and the value reported in the reference work is included in Section 3.2 in the Supporting Information. Again, we emphasize that this discrepancy does not affect the demonstration of the REXEE method since internal consistency is shown between estimates from the REXEE simulations and the benchmarks, as discussed in the next section. 

\subsubsection{Fixed-weight REXEE simulations}
\begin{figure}[ht!]
    \centering
    \includegraphics[width=\textwidth]{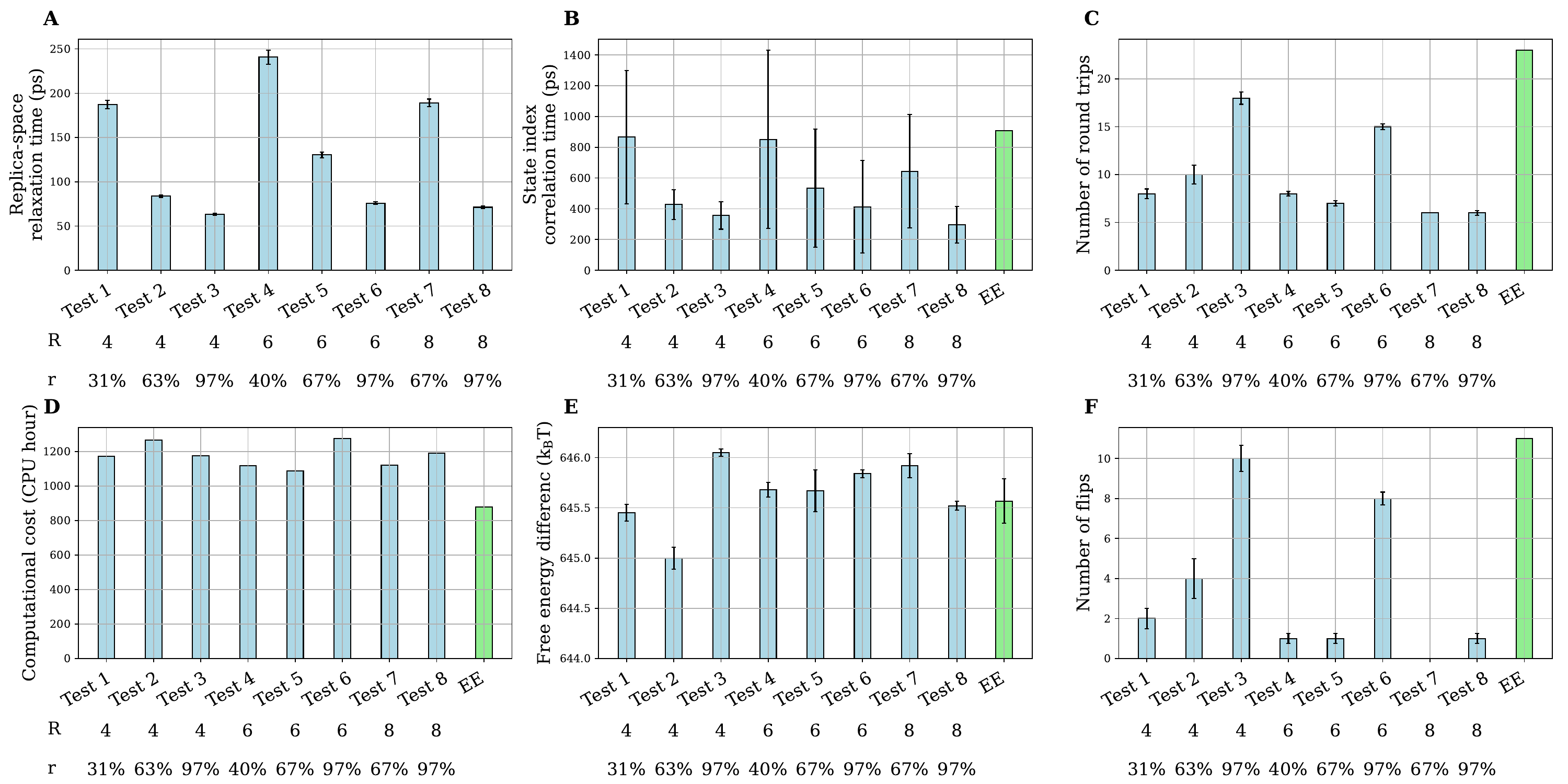}
    \caption{Results from fixed-weight REXEE simulations of the CB7-10 binding complex, , with the number of replicas ($R$) and the overlap ratio ($r$) annotated. Metrics include the (A) replica-space relaxation time, (B) state index correlation time, (C) total number of round trips in the alchemical space, (D) free energy difference $\Delta G_{\text{D}\rightarrow\text{F}}$, (E) computational cost, and (F) the number of flips between the two largest clusters of the binding complex. Results from the benchmark simulations are colored in light green. For each metric, the uncertainty was estimated using the same method used for the anthracene simulations, except that the uncertainty of the free energy benchmark from the EE simulation was calculated as the standard deviation over 3 replicates, instead of the statistical error calculated by the MBAR estimator that can occasionally be an underestimate. Overall, the trends shown in each metric of sampling are consistent with those observed in the anthracene simulations, e.g., there exists a positive correlation between the overlap ratio and state-space/replica-space sampling.}
    \label{CB7_10_sampling_speed}
\end{figure}

In panels A, B, and C in Figure \ref{CB7_10_sampling_speed}, we present the results of the three metrics assessing the sampling efficiency in the alchemical space, including the replica-space relaxation time, state index correlation time, and the total number of round trips. Overall, the exhibited trend is consistent with the observation from the anthracene REXEE simulations that a higher overlap ratio would lead to a shorter replica-space relaxation time (i.e., faster replica space sampling) and a shorter state index correlation time (i.e., faster state space sampling) and that the REXEE simulations have fewer round trips than the EE benchmark simulation. Interestingly, some tests (Tests 2, 3, 6, and 8) even show statistically shorter state index correlation time than the EE benchmark simulation (see Figure \ref{CB7_10_sampling_speed}B), which might be attributable to the fact that the exchange period ($4$ $\mathrm{ps}$) is much shorter than the intrinsic correlation time in configuration space. Another trend in the CB7-10 simulations consistent with that for the anthracene system is that all REXEE simulations, which utilized the same exchange frequency but different REXEE configurations, incurred the same level of computational cost, with reasonable overhead introduced by frequent exchanges of coordinates as compared to the benchmark EE simulation (see Figure \ref{CB7_10_sampling_speed}D).

For free energy calculations, there is some indication that the uncertainty of the free energy estimate from the EE benchmark simulation may be underestimated. This conclusion is based on observations from two additional replicates of the benchmark simulation. Specifically, we found that the standard deviation of the three replicates, $0.27$ $\mathrm{k_BT}$, was much higher than the statistical error of $0.05$ $\mathrm{k_BT}$ computed by the MBAR estimator for each replicate. This discrepancy likely lies in the fact that the data decorrelation method~\cite{chodera2016simple} occasionally underestimated the statistical inefficiency/overestimated the number of effective samples in CB7-10 trajectories. To provide a more accurate baseline for comparison with the REXEE simulations, we report the average and standard deviation over the 3 replicates of the EE benchmark simulation in Figure \ref{CB7_10_sampling_speed}E, which has a value of $-645.57\pm0.27$ $\mathrm{k_BT}$. This refined benchmark leads to a binding free energy estimate of $-21.13\pm0.27$ $\mathrm{k_BT}$, which is consistent with the estimate considering only the first replicate as reported in the previous section, i.e., $-21.33 \pm 0.07$ $\mathrm{k_BT}$. Upon comparison with the refined benchmark, we find that free energy estimates $\Delta G_{\text{D}\rightarrow\text{F}}$ from all REXEE simulations are statistically consistent with the benchmark, regardless of the replica configuration. However, we note that since the issue of underestimating the statistical inefficiency in data decorrelation also occurred in REXEE trajectories, the lower uncertainties observed in the REXEE simulations compared to the EE benchmark do not necessarily provide sufficient evidence of the superior accuracy of the REXEE method in free energy calculations. 


Finally, Figure \ref{CB7_10_sampling_speed}F shows that REXEE simulations exhibited substantially fewer transitions between the two dominant clusters corresponding to configurations of the ligand bound to different sides of the host ring molecule as compared to the EE benchmark. This difference was further elucidated in Figure S2 in the Supporting Information, where a strong correlation was observed between the number of flips and the number of round trips in the alchemical space, with a Kendall's tau correlation coefficient of 0.93. This correlation comes from the fact that the guest molecule can unbind from and rebind to the host molecule more easily in the fully decoupled state, so more switches between the two alchemical end states bring more opportunities for more configurational flips, indicating the correlation between the alchemical and configurational degrees of freedom of the binding complex. Importantly, this correlation suggests that the previously identified inherent barrier to sampling across state-set boundaries, which accounts for the reduced number of round trips in REXEE simulations, could also be responsible for the decreased number of transitions between clusters. Thus, REXEE configurations that have increased motion in state space, such as higher exchange rates and greater overlap, should generally be preferred. However, the variability in the number of flips between different REXEE configurations, and the significant uncertainty observed in the number of flips in additional replicates for some tests require further investigation, which is beyond the scope of the paper. 

\subsubsection{Weight-updating REXEE simulations}
\begin{figure}[ht!]
    \centering
    \includegraphics[width=\textwidth]{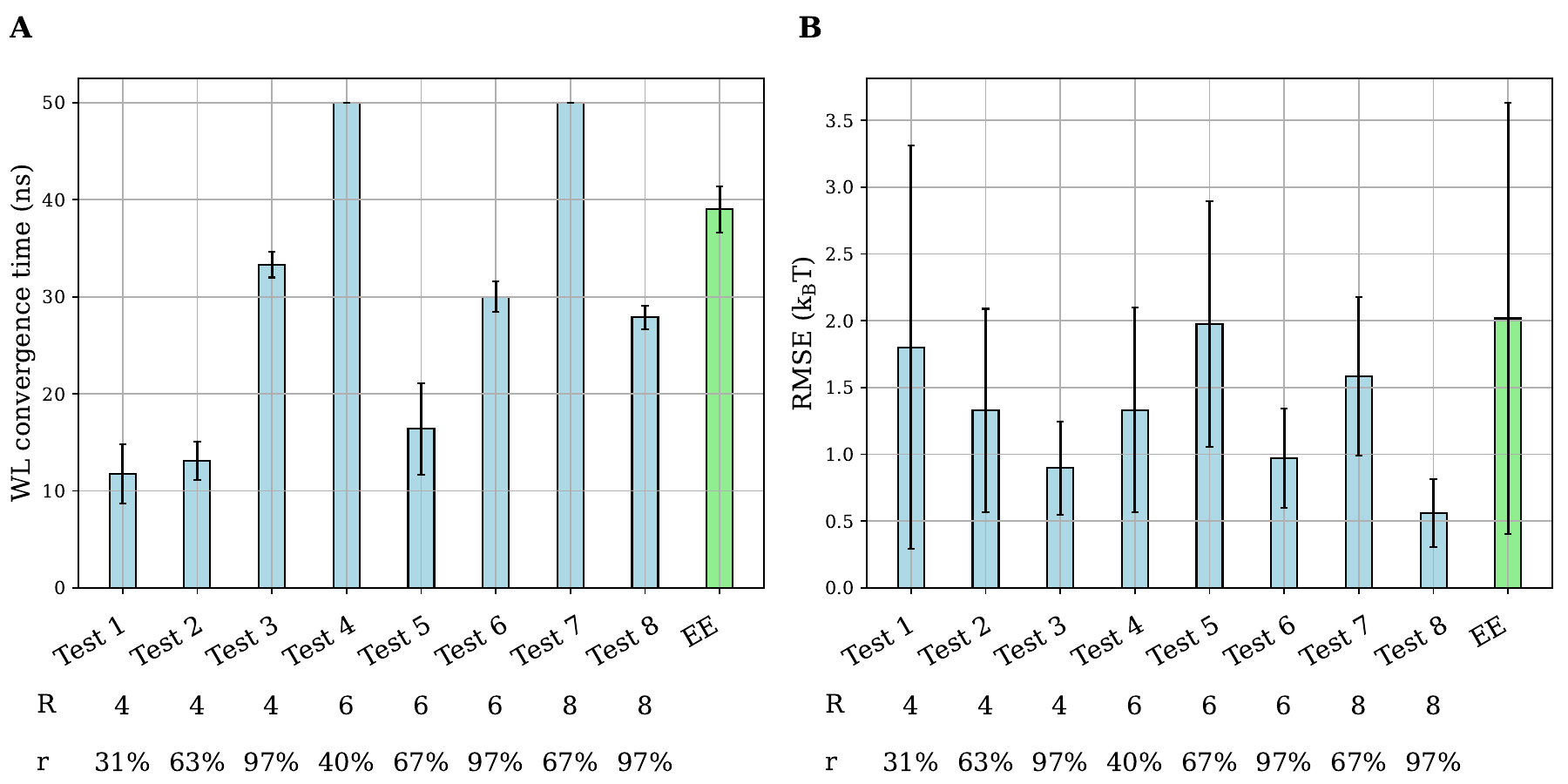} 
    \caption{(A) The Wang-Landau (WL) convergence time for all weight-updating REXEE and EE simulations, with the number of replicas ($R$) and the overlap ratio ($r$) annotated. For Tests 4 and 7, 50 ns is reported because the simulation did not converge the weights for all replicas when reaching the scheduled length of 50 ns. (B) The RMSE relative to the reference free energy profile for all weight-updating REXEE simulations and the weight-updating EE simulation. For both panels, the reported values are averages over 3 replicates, with the uncertainty being the standard deviation. The figure shows that while simulations with different overlap ratios may vary in the WL convergence time, they achieve similar RMSE values, suggesting little impact of the state set overlap on the quality of the converged weights.}
    \label{CB7_10_weight_updating}
\end{figure}

For challenging systems like binding complexes, the quality of the alchemical weights that a weight-updating simulation converges to is of importance, as the weights determine the flatness of the biased free energy surface that seeds a subsequent fixed-weight simulation. In Figure \ref{CB7_10_weight_updating}, we plotted the WL convergence times and the RMSE values relative to the reference free energy profile for all weight-updating REXEE and EE simulations of CB7-10. As a result, most tests achieved weight convergence for all replicas within 35 ns, with Tests 4 and 7 as notable exceptions, which failed to meet convergence criteria within the scheduled 50 ns simulation timeframe. Notably, those who converged within a shorter time than the weight-updating EE simulation still consumed higher computational costs, as synchronous REXEE simulations require all replicas to continue running even if some of them have converged the weights for their respective state sets. Despite these higher computational costs, Figure~\ref{CB7_10_weight_updating}B indicates that some REXEE tests could reach lower RMSE values than weight-updating EE simulations. Interestingly, while Tests 1 and 2 have statistically shorter WL convergence time than Test 3, they ended up converging to weights that have a larger deviation from the reference free energy profile. This indicates that the known issue of the Wang-Landau algorithm of getting ‘burned in’ to inaccurate free energy estimates can still persist in weight-updating REXEE simulations, and a prolonged WL convergence time may not necessarily represent a negative outcome, especially considering the pursuit of lower RMSE values for a better starting point in subsequent fixed-weight simulations. This observation also resonates with the fact that Tests 4 and 7 reached an RMSE value on par with others in spite of their inability to converge as fast as other tests. Lastly, the comparable RMSE values across all tests regardless of REXEE configurations suggest that for complex systems like CB7-10, selecting the number of replicas based on available computational resources, with a preference for an intermediate to high overlap ratio, could be a viable strategy. 

\section{Conclusions}

In this study, we proposed the method of replica exchange of expanded ensembles (REXEE), a novel generalized ensemble method that integrates the working principles of the methods of Hamiltonian replica exchange and expanded ensemble. The REXEE method decouples the number of states from the number of replicas in its algorithmic design, escaping the one-to-one correspondence inherent to the HREX method and therefore offering much greater flexibility in parameter specification---given the number of replicas decided based on the available computational resources, the REXEE method allows the number of states to be arbitrarily chosen. In addition, the parallelizability of the REXEE method opens the door to wider applications compared to the EE method, especially one-shot free energy calculations in multi-topology contexts such as serial mutations or scaffold-hopping transformations. All necessary algorithms for running and analyzing REXEE simulations have been encapsulated in the Python package \verb|ensemble_md|, which is a flexible and easily extensible wrapper that not only eliminates the need to modify the source code of GROMACS, but also provides convenient automation and management of REXEE simulations.

With this implementation, we evaluated the effectiveness of the REXEE method with various setups, focusing on the calculation of the solvation free energy of anthracene and the binding free energy of the CB7-10 complex. As a result, we confirmed that the REXEE method did not compromise its ability to accurately compute free energy differences while offering enhanced flexibility and parallelizability. In addition, we observed that increasing the exchange frequency or overlap ratio in fixed-weight REXEE simulations enhances sampling/mixing in both the replica and state spaces without introducing exorbitant overhead as compared to EE and HREX simulations. In weight-updating simulations, we demonstrated that the exchange frequency and REXEE configuration have little impact on the WL convergence time and the RMSE value relative to the reference free energy profile. Moreover, we concluded that weight combination schemes could actually introduce additional errors in the alchemical weights, leading to longer WL convergence times without necessarily improving the accuracy of the weights. Lastly, while weight-updating REXEE simulations may incur slightly higher computational costs than weight-updating EE simulations, our results indicate that the REXEE method holds the potential to converge to more accurate weights, perhaps by incorporating additional initial configurations, promising a better starting point to seed subsequent fixed-weight simulations.

Importantly, the current REXEE implementation could be further enhanced by incorporating asynchronous parallelization schemes, which permit less constrained communications between a larger count of loosely coupled processors, making it more adaptable to environments like cloud computing, compared to the currently implemented synchronous REXEE method. In the weight-updating phase, asynchronous REXEE can save computational power by dropping replicas that have converged the corresponding alchemical weights. This contrasts with synchronous parallelization schemes, where the termination of a single replica halts the entire simulation ensemble. Moreover, asynchronous REXEE provides an elevated level of flexibility in parameter specification compared to synchronous REXEE. It not only accommodates heterogeneous parameter configurations, such as different numbers of states per replica or state shifts for different replicas, but also allows for adaptive changes to the parameters in response to data collected, such as changing the number of states per replica or even the number of replicas to optimize the simulations' performance. Lastly, the development of the package \verb|ensemble_md| and the foundational principles of the REXEE method can be integrated with our recent endeavors in alchemical metadynamics by having multiple walkers sample the joint space of alchemical variable and configurational collective variables, with each walker exploring different alchemical state sets. This can potentially result in the development of new simulation approaches like the ensemble of alchemical metadynamics, promising enhanced flexibility, parallelizability, and configurational sampling for a wider range of systems.

\section*{Author Contributions}
W.-T.H. and M.R.S. primarily conceptualized the project and designed the methodology. W.-T.H. implemented the sampling method in the Python package \verb|ensemble_md|. Experiments were performed and analyzed by W.-T.H., with contributions from M.R.S. for validation. W.-T.H wrote the original manuscript draft; editing and review of the manuscript were done by M.R.S. M.R.S. supervised the project and obtained resources.

\begin{acknowledgement}
This study was supported by the grant OAC-1835720 from the National Science Foundation and grant R01GM123296 from the National Institutes of General Medical Sciences. The computational work done in this publication used resources provided from the Extreme Science and Engineering Discovery Environment (XSEDE), which is supported by National Science Foundation grant number ACI-1548562. Specifically, it used the Bridges-2 system, which is supported by NSF ACI-1928147, located at the Pittsburgh Supercomputing Center (PSC).  We thank Anika Friedman for proofreading and comments on the manuscript.
\end{acknowledgement}

\newpage
\begin{figure}[ht]
    \centering
    \includegraphics[width=\textwidth]{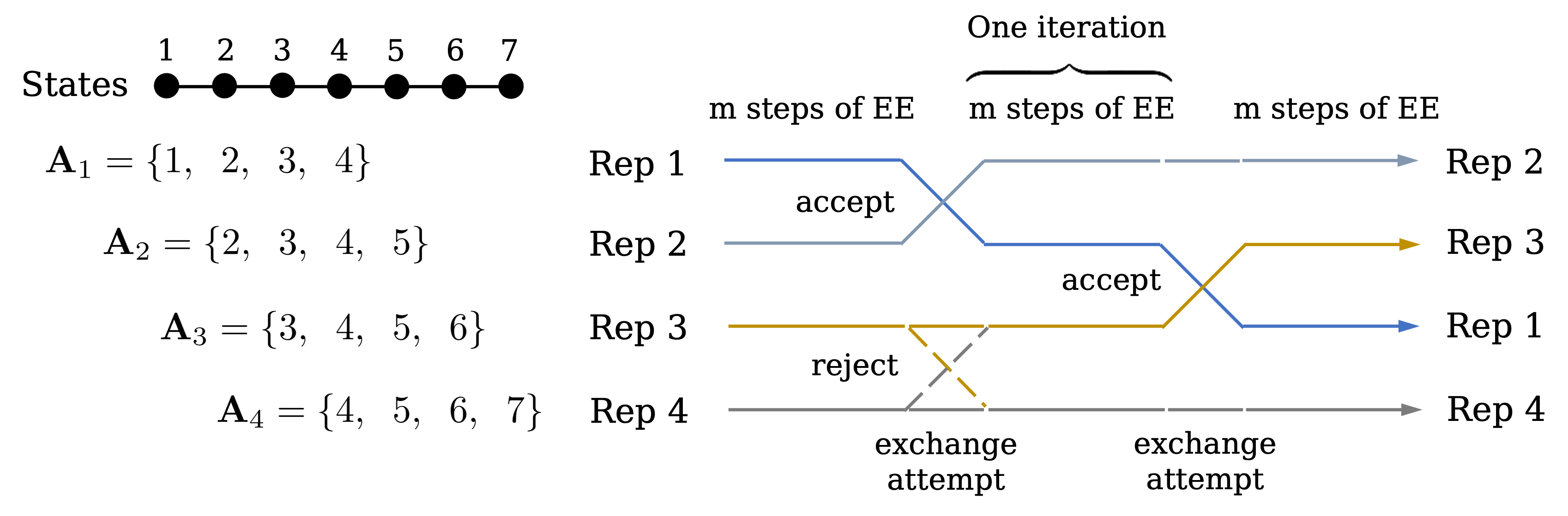}   
    \caption{TOC}
\end{figure}


\clearpage
\input{main.bbl}


\end{document}


\section{1. Enumerating the space of configurations of REXEE simulations}
Configurational parameters of a REXEE simulation include $N$ (the number of states in total), $n_s$ (the number of states in each replica), $R$ (the number of replicas), and $\phi$ (the state shift between adjacent state sets). Given their correlation with the sampling efficiency of REXEE simulations, it is crucial to set these parameters carefully. Here, we discuss an efficient way to explore the parameter space of homogeneous REXEE simulations with one-dimensional sequential alchemical intermediates given a specified value of $N$, which is typically decided to ensure sufficient overlap between neighboring states. 

In the case where all replicas sample different but overlapping sets of states in a REXEE simulation, all configurational parameters must be integers and have a physical minimum, i.e., $N\geq 3$, $R \geq 2$, $n_s \geq 2$, and $\phi \geq 1$.  To ensure a non-zero state overlap between neighboring state sets, it is additionally required that $n-s\geq 1$. All these configurational parameters are related by the following expression 
\begin{equation}
    n_s + (R-1) \phi=N \label{config_EEXE}
\end{equation}Notably, there often exists a moderately low number of choices for the number of replicas $R$, which is preferably specified as a factor of the number of available cores for efficient parallelization. In the case where we loop over these limited choices of $R$, Equation \ref{config_EEXE} becomes a linear Diophantine equation and can be solved by the following lemma. 

\begin{lemma}
Let $a_1, ..., a_n \in \mathbb{Z}$, a general linear Diophantine equation in the form $a_1x_1 + ... + a_nx_n=c$ have solutions if and only if $\gcd(a_1, ..., a_n)|c$. If $\gcd(a_1, ..., a_n)|c$, there are infinitely many solutions that can be expressed with $n-1$ parameters. 
\end{lemma}

Specifically, when $N$ is known and $R$ is specified, Equation \ref{config_EEXE} reduces to a two-variable Diophantine equation, which fits in the following simplified lemma: 

\begin{lemma}
Let $a, b, c \in \mathbb{Z}$, if $\gcd(a, b)=c$, the Diophantine equation $ax + by = c$ has infinitely many solutions of the form \begin{equation}
  \begin{cases} 
  x=x_0+\left(\frac{b}{\gcd(a, b)}\right)t \\       
  y=y_0 - \left(\frac{a}{\gcd(a, b)}\right)t\\
  \end{cases}
\end{equation}where $(x_0, y_0)$ is a particular solution and $t\in\mathbb{Z}$. 
\end{lemma}

In our case, with $a=1$ and $b=R-1$ and $\gcd(a, b)=\gcd(1, r-1)=1$, we have \begin{equation}
  \begin{cases} 
  x=x_0 + bt \\       
  y=y_0 - at\\
  \end{cases} \Rightarrow \begin{cases} 
  n_s=n_{s, 0} + (R-1)t \\       
  \phi=\phi_0 - t\\
  \end{cases}
\end{equation}For any given $r$ and $N$ values, there should always be a solution of $\phi=\phi_0=1$, and $n_s=n_{s, 0}=N+1-R$, which can serve as the particular solution that leads us to 
\begin{equation}
  \begin{cases} 
  n_s=N+(R-1)(t-1) \\       
  \phi=1 - t\\
\end{cases}
\end{equation}
To identify the bounds of $t$, we examine each of the constraints described above. Specifically, given that $s\geq 1$, we have
\begin{equation}
    s=1-t \geq 1 \Rightarrow t\leq 0
\end{equation}With $n\geq2$, we have 
\begin{equation}
    n=N+(r-1)(t-1) \geq2 \Rightarrow t\geq \frac{r-N+1}{r-1} \label{n_geq_2}
\end{equation}And finally, the condition $n-s\geq1$ requires the following
\begin{equation}
    N + (R-1)(t-1) -(1-t) = N + R(t-1)\geq 1 \Rightarrow t\geq \frac{R-N+1}{R}  \label{n-s_geq_1 }
\end{equation}Given that $\frac{R-N+1}{R-1}\leq\frac{R-N+1}{R}\leq 0$, Equations \ref{n_geq_2} and \ref{n-s_geq_1 } collectively gives
\begin{equation}
    \frac{R-N+1}{R}\leq t\leq0
\end{equation}This derivation leads to the following conclusion:

\begin{thm}
Consider a homogeneous REXEE simulation that collectively samples $N$ alchemical intermediate states in total sequentially defined in one-dimensional grids. Given that the ensemble is composed of $R$ replicas, the number of states for each replica $n_s$ and the state shift $\phi$ can be expressed using the following parametric equation: \begin{equation}
  \begin{cases} 
  n_s=N + (R-1)(t-1) \\       
  \phi=1 - t\\
  \end{cases}\label{conclusion}
\end{equation} where $t \in \mathbb{Z}$ and $\frac{R-N+1}{R} \leq t \leq 0$. 
\end{thm}

Programmatically, this provides a systematic way to find all the possible $(N, R, n_s, \phi)$ combinations when the value of $N$ is specified. Specifically, one only needs to loop over the $R$ values of interest, and for each $R$ value, substitute all integers $t$ in the interval $[\frac{R-N+1}{R}, 0]$ to Equation \ref{conclusion} to get the values of $n_s$ and $\phi$. Conveniently, it is trivial to apply constraints to any configurational parameter. For example, in the case where a maximum overlap ratio of 50\% is desired ($(n_s-\phi)/n_s \leq 0.5$), we get $t\leq\frac{R-N+1}{R+1}$, which allows us to further narrow down the range of $t$ as follows:
\begin{equation}
\frac{R-N+1}{R}\leq t \leq \frac{R-N+1}{R+1}
\end{equation}
This method has been implemented in the CLI \verb|explore_REXEE| in the package \verb|ensemble_md|, with which we can effortlessly explore the parameter space of REXEE simulations. For example, Figure S1A shows the number of possible $(N, R, n_s, \phi)$ combinations roughly increases linearly with the total number of states. More importantly, Figure S1B shows that for any REXEE simulation having more than 3 alchemical states in total ($N > 3$), there exists at least one solution of $(R, n_s, \phi)$ for any $R$ satisfying $2\leq R\leq N-1$, which underpins that the REXEE method has much higher flexibility in specifying the value of $N$ than the HREX method.

\begin{figure}[H]
    \centering
    \includegraphics[width=\textwidth]{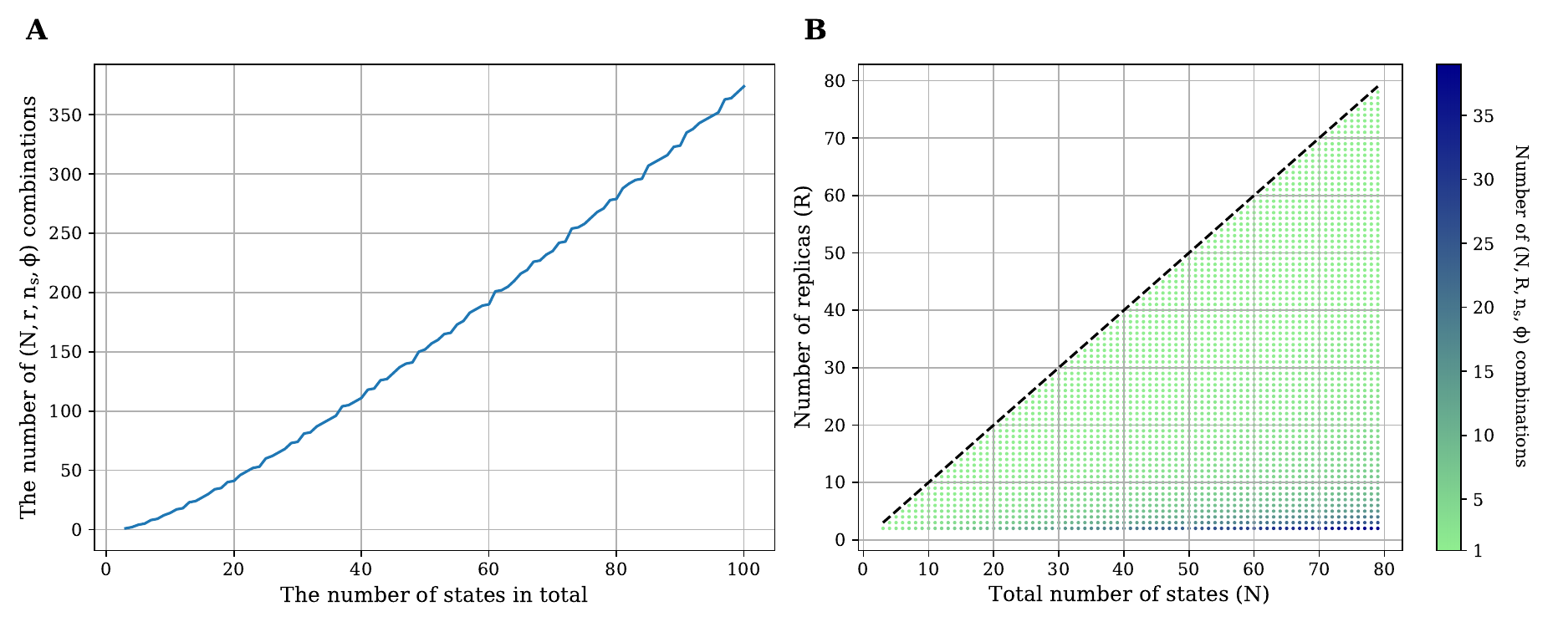}   
    \caption{(A) The number of possible $(N, R, n_s, \phi)$ combinations as a function of the total number of states. (B) The parameter landscape of REXEE simulations with dots colored with the number of the $(N, R, n_s, \phi)$ combinations. The black dashed line is a 45-degree line, under which shows the region for valid $N$ value for each $R$ value given the natural constraint that the number of replicas $R$ should be smaller than the number of states $N$. Since the minimum of the color bar is 1, the fact that all points (i.e., $(N, R)$ pairs) under the dashed line are colored indicates that for any given number of replicas, there exist at least one $(N, R, n_s, \phi)$ combination.}
\end{figure}


\section{2. Derivation of the acceptance ratio for swapping EE replicas in a REXEE simulation}
In this section, we will first define thermodynamic states in REXEE simulations, and then derive the acceptance ratio that enforces the detailed balance condition at the inter-replica level assuming symmetric proposal probabilities. Note that all subsequent derivations apply for REXEE simulations with any replica configuration. 

\subsubsection{2.1. Thermodynamic states in a REXEE simulation}
To make the derivation as general as possible, we define the reduced potential $u(x)$ of a system as \begin{equation}
    u(x) = \beta \left[H(x) + pV(x) + \sum_k \mu_k n_k(x) + \cdots \right]
\end{equation} where $\beta=(k_\text{B}T)^{-1}$ is the inverse temperature and $x \in \Omega$ denotes the coordinates of the system, with $\Omega$ being the accessible configurational space. This reduced potential not only includes the Hamiltonian $H(x)$ of the system, but also encompasses energy terms associated with the change of different physical variables, including the volume $V(x)$ (in a constant-pressure ensemble with a pressure $p$) and $n_k(x)$, the number of molecules of each of $k=1, ..., M$ components of the system, and the corresponding chemical potential $\mu_k$ (in a (semi-)grand ensemble).

Here, we define $X=\left(..., x^i_m, ..., x^j_n, ...\right)$ of length $R$ as a ``state'' in the generalized ensemble sampled by a REXEE simulation, where $x^i_m\equiv \left(x_i, A_m\right)$, meaning that the $i$-th replica samples the $m$-th state set with the coordinates $x_i$. Then, the joint probability of such a replica sampling the state $s_i$ in the state set $A_m$ (with $s_i \in A_m$) is given by\begin{equation}
    \pi\left(x^i_m\right) =\dfrac{\exp\left(-u_{s_i}(x_i) + g^m_{s_i}\right)}{\displaystyle\sum\limits_{s\in A_m}\left(\displaystyle\int_{\Omega} \exp\left(-u_s(x) + g^m_s\right)dx\right)}
\end{equation} where $u_{s_i}$ is the reduced potential of state $s_i$ and $g^m_{s_i}$ is the weight (in units of $k_\text{B}T$) of state $s_i$ in the $m$-th state set. We note that the absence of the state set index as the superscript for the reduced potential reflects the fact that the functional form of the reduced potential of state $s_i$ should be the same across all replicas and only vary with different intermediate states $s_i$. Additionally, the same alchemical state in different state sets may carry different alchemical weights, namely, it is possible that $g^m_{s_i}\neq g^n_{s_j}$ even if $s_i = s_j$. This situation can arise when a REXEE simulation consists of fixed-weight EE simulations (termed fixed-weight REXEE) having different weights specified for the same state $s$ across different state sets. It can also occur in a REXEE simulation composed of weight-updating EE simulations (termed weight-updating REXEE). In the latter case, the cumulative alchemical weights depend on the visitation of the alchemical states by the Monte Carlo moves during a weight-updating process, which introduces stochasticity that causes the difference between $g^m_{s_i}$ and $g^n_{s_j}$ even when $s_i=s_j$.

\subsubsection{2.2. Acceptance ratio}
With $X=\left(..., x^i_m, ..., x^j_n, ...\right)$, we consider replicas $i$ and $j$ that sample the state sets $A_m$ and $A_n$, respectively. To swap replicas $i$ and $j$ (which is equivalent to swapping the state sets $A_m$ and $A_n$), the state sampled by replica $i$ at the moment must fall within the state set $A_n$ that is to be swapped, and vice versa. In this case, we call that these replicas $i$ and $j$ are ``swappable''. The list of swappable pairs $\mathcal{S}$ can be defined as the set of replica pairs as follows:\begin{equation}
    \mathcal{S} = \{(i, j)|s_i \in A_n \text{ and } s_j \in A_m, i\neq j\}
\end{equation}
Exchanging replicas $i$ and $j$ results in a transition as follows:
\begin{equation}
    X=\left(..., x^i_{m}, ..., x^j_{n}, ...\right) \rightarrow X' = \left(..., x^j_{m}, ..., x^i_{n}, ...\right)\label{swappable}
\end{equation} which necessarily introduces a new permutation $f'$ such that $f'(j)=m$ and $f'(i)=n$. 

In this setup, the following equation must be fulfilled by each replica $i$ to achieve detailed balance at the inter-replica level:\begin{equation}
    \alpha(X'|X)\pi(X)P_{\text{acc}}(X'|X) = \alpha(X|X')\pi(X')P_{\text{acc}}(X|X')\label{detailed_balance}
\end{equation}
where $\alpha\left(X'|X\right)= \alpha\left(x^j_{m}, x^i_{n} | x^i_{m}, x^j_{n}\right)$ is the probability of proposing a transition from states $X$ to $X'$ , $P_{\text{acc}}\left(X'|X\right)= P_{\text{acc}}\left(x^j_{m}, x^i_{n} | x^i_{m}, x^j_{n}\right)$ is the probability that such a transition is accepted, and $\pi(X)$ is the probability that the system is at state $X$. As will be discussed in the next section, most of the straightforward proposal schemes have symmetric proposal probabilities, i.e., $\alpha(X'|X)=\alpha(X|X')$, in which case Equation \ref{detailed_balance} is reduced to \begin{equation}
    \pi(X)P_{\text{acc}}(X'|X) = \pi(X')P_{\text{acc}}(X|X')
\end{equation}
Given the thermodynamic state $X$ defined above, $\pi(X)$ can be expressed as follows:\begin{equation}
    \begin{aligned}
    \pi(X) &= \cdots \pi\left(x^i_{m}\right) \cdots \pi\left(x^{j}_{n}\right) \cdots = \prod_{k=1}^R \pi\left(x^k_{f(k)}\right)\\ &= \prod_{k=1}^R \left(\dfrac{\exp\left(-u_{s_k}(x_k) + g^{f(k)}_{s_k}\right)}{\displaystyle\sum\limits_{s\in A_{f(k)}}\left(\displaystyle\int_{\Omega} \exp\left(-u_s(x) + g^{f(k)}_s\right)dx\right)}\right) \label{pi}
\end{aligned}
\end{equation}

Then $\pi\left(X'\right)$ will be given by \begin{equation}
\begin{aligned}
    \pi\left(X'\right) &= \cdots \pi\left(x^i_n\right) \cdots \pi\left(x^j_m\right) \cdots = \prod_{k=1}^R \pi\left(x^k_{f'(k)}\right)  
    \end{aligned}
\end{equation}
Canceling the stationary probabilities of the non-swapping replicas, we have \begin{equation}
\begin{aligned}
    \dfrac{P_{\text{acc}}(X'|X)}{P_{\text{acc}}(X|X')} &= \dfrac{\pi(X')}{\pi(X)} = \dfrac{\pi\left(x^j_{m}\right)\pi\left(x^i_{n}\right)}{\pi\left(x^i_{m}\right)\pi\left(x^j_{n}\right)} \\ &= \dfrac{\left(\dfrac{\exp\left(-u_{s_i}(x_j) + g^{m}_{s_i}\right)}{\bcancel{\displaystyle\sum\limits_{s\in A_{m}}\left(\displaystyle\int_{\Omega} \exp\left(-u_s(x) + g^{m}_s\right)dx\right)}}\right)\left(\dfrac{\exp\left(-u_{s_j}(x_i) + g^{n}_{s_j}\right)}{\cancel{\displaystyle\sum\limits_{s\in A_{n}}\left(\displaystyle\int_{\Omega} \exp\left(-u_{s}(x) + g^{n}_s\right)dx\right)}}\right)}{\left(\dfrac{\exp\left(-u_{s_i}(x_i) + g^{m}_{s_i}\right)}{\bcancel{\displaystyle\sum\limits_{s\in A_{m}}\left(\displaystyle\int_{\Omega} \exp\left(-u_s(x) + g^{m}_s\right)dx\right)}}\right)\left(\dfrac{\exp\left(-u_{s_j}(x_j) + g^{n}_{s_j}\right)}{\cancel{\displaystyle\sum\limits_{s\in A_{n}}\left(\displaystyle\int_{\Omega} \exp\left(-u_s(x) + g^{n}_s\right)dx\right)}}\right)} \\
\end{aligned}
\end{equation} This yields \begin{equation}
\begin{aligned}
    \dfrac{P_{\text{acc}}(X'|X)}{P_{\text{acc}}(X|X')}  &= \exp\left[\left(-u_{s_i}(x_j) + \cancel{g^m_{s_i}}\right)- \left(-u_{s_i}(x_i) + \cancel{g^m_{s_i}}\right)+\left(-u_{s_j}(x_i) + \bcancel{g^n_{s_j}}\right)-\left(-u_{s_j}(x_j) + \bcancel{g^n_{s_j}}\right)\right] \\&=\exp\left[\left(-u_{s_i}(x_j)\right)- \left(-u_{s_i}(x_i)\right)+\left(-u_{s_j}(x_i)\right)-\left(-u_{s_j}(x_j)\right)\right] \\&=\exp(-\Delta)
\end{aligned}
\end{equation} where 
\begin{equation}
    \Delta = \left(u_{s_i}(x_j) + u_{s_j}(x_i) \right)-\left(u_{s_i}(x_i)+u_{s_j}(x_j)\right)\label{Delta}
\end{equation}
Notably, this expression of $\Delta$ is of the same form as that in the HREX method. Using the Metropolis criterion, the acceptance ratio $P_{\text{acc}}(X\rightarrow X')$ can be written as \begin{equation}
  P_{\text{acc}}(X \rightarrow X') = 
  \begin{cases} 
    \begin{aligned}
      &1 &, \text{if } \Delta \leq 0 \\
      \exp(&-\Delta) &, \text{if } \Delta >0
    \end{aligned}
  \end{cases}
  \label{p_acc}
\end{equation}

\section{3. Comparison between our free energy estimates and references reported in previous studies}
\subsection{3.1. Solvation free energy of anthracene}
As mentioned in the main text, the free energy benchmarks from EE and HREX simulations were 3.502 $\pm$ 0.178 $\mathrm{k_BT}$ and 3.411 $\pm$ 0.067 $k_{\text{B}}T$, respectively. These two values, however, do not agree with the value reported in the work by Paliwal et al., ~\cite{paliwal2011benchmark} despite the same system topology and mostly the same simulation parameters. Specifically, in their work, the solvation free energy of anthracene was reported as 2.611 $\pm$ 0.006 $k_\text{B}T$, which was computed using MBAR~\cite{shirts2008statistically} from 450 ns of direct sampling of 15 alchemical states in an NPT ensemble. Since the impact of different sampling approaches, simulation ensembles and alchemical paths are expected to be small, and all simulations performed in this study are consistent, we attribute the discrepancy in the free energy estimates mainly to different versions of GROMACS and data decorrelation methods. Specifically, simulations in the work by Paliwal et al. were done with GROMACS 4.0.4, a version more than a decade ago that could make the direct comparison difficult. Importantly, the disagreement in the free energy estimates between our benchmarks and the previously published result does not necessarily affect our demonstration of the REXEE method, as long as internal consistency is achieved between the results from the REXEE simulations and the benchmark values obtained from EE and HREX simulations. 

\subsection{3.2. Binding free energy of CB7-10 binding complex}
In the main text, we reported the benchmark of the binding free energy calculated from the EE simulation as $\Delta G^{\circ}_{\text{bind}}$ as $-21.33 \pm 0.07$ $\mathrm{k_BT}$. As a reference, the binding free energy of CB7-10 reported in the work by Monroe et al.~\cite{monroe2014converging} was $-8.34 \pm 0.32$ $\mathrm{kcal/mol}$, or $-13.99 \pm 0.54$ $\mathrm{k_BT}$. While there exists a discrepancy between our estimate and the reference, we note that an incorrect analytical expression of the correction term was used in the reference work. Specifically, the reference work adopted a non-zero reference distance for the distance restraint between the host and guest molecules, but calculated the correction term using an equation similar to that for the case where a zero reference distance is used. Additionally, the equation itself was incorrectly derived even assuming a zero reference distance, which yielded the correction term as $3.98$ $\mathrm{kcal/mol}$, or $6.676$ $\mathrm{k_B T}$ and already accounts for a difference of $4.416$ $\mathrm{k_B T}$ as compared to our correction term calculated using Equation 16 in the main text. Assuming the same reference distance (i.e., $r_0=0.428$ $\mathrm{nm}$) used in the work by Monroe et al.~\cite{monroe2014converging} and using WebPlotDigitizer~\cite{Rohatgi2024} to extract the free energy differences $\Delta G_{\text{A}\rightarrow \text{B}}$ and $\Delta G_{\text{D}\rightarrow \text{F}}$ respectively from Figures S3 and S4 in the reference work, we estimated the reference value that should have been reported as $-19.66 \pm 0.40$ $\mathrm{kT}$, which is reasonably close to our estimate. Notably, the value of $\Delta G_{\text{A}\rightarrow\text{B}}$ extracted from Figure S3 of the reference work is $622.33 \pm 0.31$ $\mathrm{k_B T}$, which is statistically consistent with our estimate from the solvent simulation. This shows that the remaining difference between our estimate and the refined reference primarily comes from the complex simulation, which we attribute to the fact that the reference simulation was carried out in a different ensemble (NPT instead of NVT ensemble) using a GROMACS version 10 years earlier than the version we used. Again, we emphasize that the discrepancy between our estimate and the reference should not affect the demonstration of the REXEE method as long as internal consistency is achieved.

\section{4. Supplementary Figures}

\begin{figure}[H]
    \centering
    \includegraphics[width=\textwidth]{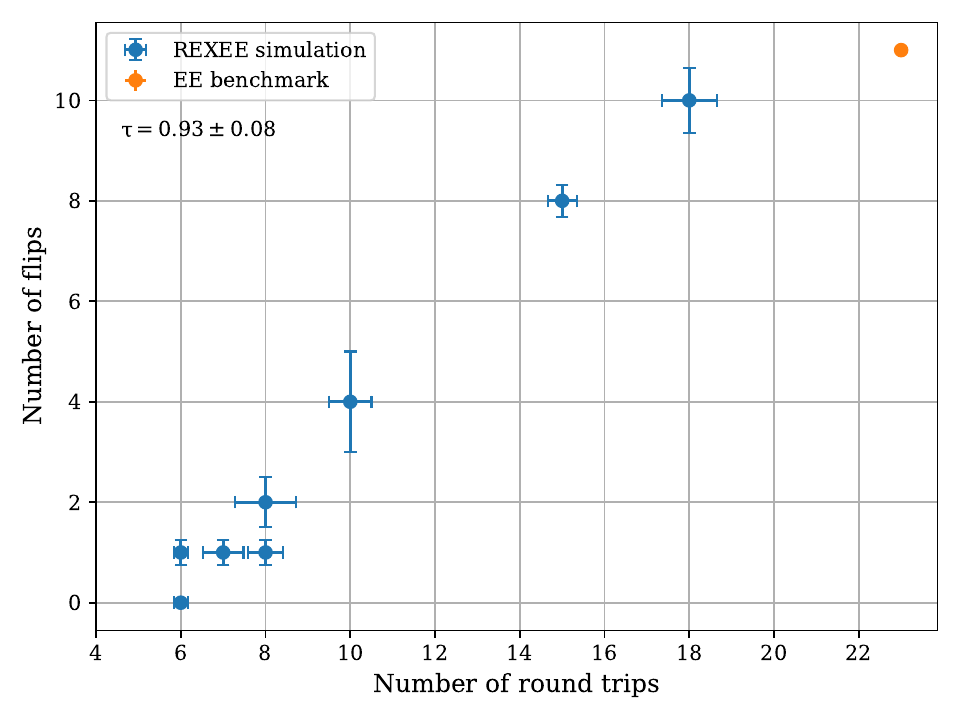}
    \caption{The parity plot of the total number of flips against the number of round trips in the alchemical space, with Kendall’s tau correlation coefficient ($\tau$) annotated.}
\end{figure}

\input{refs.bbl}

%% file: main.bbl
\providecommand{\latin}[1]{#1}
\makeatletter
\providecommand{\doi}
  {\begingroup\let\do\@makeother\dospecials
  \catcode`\{=1 \catcode`\}=2 \doi@aux}
\providecommand{\doi@aux}[1]{\endgroup\texttt{#1}}
\makeatother
\providecommand*\mcitethebibliography{\thebibliography}
\csname @ifundefined\endcsname{endmcitethebibliography}
  {\let\endmcitethebibliography\endthebibliography}{}

%% file: refs.bbl
\providecommand{\latin}[1]{#1}
\makeatletter
\providecommand{\doi}
  {\begingroup\let\do\@makeother\dospecials
  \catcode`\{=1 \catcode`\}=2 \doi@aux}
\providecommand{\doi@aux}[1]{\endgroup\texttt{#1}}
\makeatother
\providecommand*\mcitethebibliography{\thebibliography}
\csname @ifundefined\endcsname{endmcitethebibliography}
  {\let\endmcitethebibliography\endthebibliography}{}

%% file: main.bbl
\begin{mcitethebibliography}{77}
\providecommand*\natexlab[1]{#1}
\providecommand*\mciteSetBstSublistMode[1]{}
\providecommand*\mciteSetBstMaxWidthForm[2]{}
\providecommand*\mciteBstWouldAddEndPuncttrue
  {\def\EndOfBibitem{\unskip.}}
\providecommand*\mciteBstWouldAddEndPunctfalse
  {\let\EndOfBibitem\relax}
\providecommand*\mciteSetBstMidEndSepPunct[3]{}
\providecommand*\mciteSetBstSublistLabelBeginEnd[3]{}
\providecommand*\EndOfBibitem{}
\mciteSetBstSublistMode{f}
\mciteSetBstMaxWidthForm{subitem}{(\alph{mcitesubitemcount})}
\mciteSetBstSublistLabelBeginEnd
  {\mcitemaxwidthsubitemform\space}
  {\relax}
  {\relax}

\bibitem[Padua(2011)]{padua2011encyclopedia}
Padua,~D. \emph{Encyclopedia of parallel computing}; Springer Science \&
  Business Media, 2011\relax
\mciteBstWouldAddEndPuncttrue
\mciteSetBstMidEndSepPunct{\mcitedefaultmidpunct}
{\mcitedefaultendpunct}{\mcitedefaultseppunct}\relax
\EndOfBibitem
\bibitem[H{\'e}nin \latin{et~al.}(2022)H{\'e}nin, Leli{\`e}vre, Shirts,
  Valsson, and Delemotte]{henin2022enhanced}
H{\'e}nin,~J.; Leli{\`e}vre,~T.; Shirts,~M.~R.; Valsson,~O.; Delemotte,~L.
  Enhanced sampling methods for molecular dynamics simulations. \emph{arXiv
  preprint arXiv:2202.04164} \textbf{2022}, \relax
\mciteBstWouldAddEndPunctfalse
\mciteSetBstMidEndSepPunct{\mcitedefaultmidpunct}
{}{\mcitedefaultseppunct}\relax
\EndOfBibitem
\bibitem[Torrie and Valleau(1977)Torrie, and Valleau]{umbrella}
Torrie,~G.~M.; Valleau,~J.~P. Nonphysical sampling distributions in Monte Carlo
  free-energy estimation: Umbrella sampling. \emph{J. Comput. Phys.}
  \textbf{1977}, \emph{23}, 187--199\relax
\mciteBstWouldAddEndPuncttrue
\mciteSetBstMidEndSepPunct{\mcitedefaultmidpunct}
{\mcitedefaultendpunct}{\mcitedefaultseppunct}\relax
\EndOfBibitem
\bibitem[Laio and Parrinello(2002)Laio, and Parrinello]{metad}
Laio,~A.; Parrinello,~M. Escaping free-energy minima. \emph{Proc. Natl. Acad.
  Sci.} \textbf{2002}, \emph{99}, 12562--12566\relax
\mciteBstWouldAddEndPuncttrue
\mciteSetBstMidEndSepPunct{\mcitedefaultmidpunct}
{\mcitedefaultendpunct}{\mcitedefaultseppunct}\relax
\EndOfBibitem
\bibitem[Darve \latin{et~al.}(2008)Darve, Rodr{\'\i}guez-G{\'o}mez, and
  Pohorille]{ABF}
Darve,~E.; Rodr{\'\i}guez-G{\'o}mez,~D.; Pohorille,~A. Adaptive biasing force
  method for scalar and vector free energy calculations. \emph{J. Chem. Phys.}
  \textbf{2008}, \emph{128}, 144120\relax
\mciteBstWouldAddEndPuncttrue
\mciteSetBstMidEndSepPunct{\mcitedefaultmidpunct}
{\mcitedefaultendpunct}{\mcitedefaultseppunct}\relax
\EndOfBibitem
\bibitem[Invernizzi(2021)]{invernizzi2021opes}
Invernizzi,~M. OPES: On-the-fly probability enhanced sampling method.
  \emph{arXiv preprint arXiv:2101.06991} \textbf{2021}, \relax
\mciteBstWouldAddEndPunctfalse
\mciteSetBstMidEndSepPunct{\mcitedefaultmidpunct}
{}{\mcitedefaultseppunct}\relax
\EndOfBibitem
\bibitem[Piana and Laio(2007)Piana, and Laio]{var2}
Piana,~S.; Laio,~A. A bias-exchange approach to protein folding. \emph{J. Phys.
  Chem. B} \textbf{2007}, \emph{111}, 4553--4559\relax
\mciteBstWouldAddEndPuncttrue
\mciteSetBstMidEndSepPunct{\mcitedefaultmidpunct}
{\mcitedefaultendpunct}{\mcitedefaultseppunct}\relax
\EndOfBibitem
\bibitem[Invernizzi and Parrinello(2022)Invernizzi, and
  Parrinello]{invernizzi2022exploration}
Invernizzi,~M.; Parrinello,~M. Exploration vs convergence speed in
  adaptive-bias enhanced sampling. \emph{Journal of Chemical Theory and
  Computation} \textbf{2022}, \emph{18}, 3988--3996\relax
\mciteBstWouldAddEndPuncttrue
\mciteSetBstMidEndSepPunct{\mcitedefaultmidpunct}
{\mcitedefaultendpunct}{\mcitedefaultseppunct}\relax
\EndOfBibitem
\bibitem[Dama \latin{et~al.}(2014)Dama, Rotskoff, Parrinello, and
  Voth]{dama2014transition}
Dama,~J.~F.; Rotskoff,~G.; Parrinello,~M.; Voth,~G.~A. Transition-tempered
  metadynamics: robust, convergent metadynamics via on-the-fly transition
  barrier estimation. \emph{Journal of Chemical Theory and Computation}
  \textbf{2014}, \emph{10}, 3626--3633\relax
\mciteBstWouldAddEndPuncttrue
\mciteSetBstMidEndSepPunct{\mcitedefaultmidpunct}
{\mcitedefaultendpunct}{\mcitedefaultseppunct}\relax
\EndOfBibitem
\bibitem[Chodera and Shirts(2011)Chodera, and Shirts]{chodera2011replica}
Chodera,~J.~D.; Shirts,~M.~R. Replica exchange and expanded ensemble
  simulations as Gibbs sampling: Simple improvements for enhanced mixing.
  \emph{J. Chem. Phys.} \textbf{2011}, \emph{135}, 194110\relax
\mciteBstWouldAddEndPuncttrue
\mciteSetBstMidEndSepPunct{\mcitedefaultmidpunct}
{\mcitedefaultendpunct}{\mcitedefaultseppunct}\relax
\EndOfBibitem
\bibitem[Geman and Geman(1984)Geman, and Geman]{geman1984stochastic}
Geman,~S.; Geman,~D. Stochastic relaxation, Gibbs distributions, and the
  Bayesian restoration of images. \emph{IEEE Trans. Pattern Anal. Mach.
  Intell.} \textbf{1984}, 721--741\relax
\mciteBstWouldAddEndPuncttrue
\mciteSetBstMidEndSepPunct{\mcitedefaultmidpunct}
{\mcitedefaultendpunct}{\mcitedefaultseppunct}\relax
\EndOfBibitem
\bibitem[Hastings(1970)]{hastings1970monte}
Hastings,~W.~K. Monte Carlo sampling methods using Markov chains and their
  applications. \emph{Biometrika} \textbf{1970}, \relax
\mciteBstWouldAddEndPunctfalse
\mciteSetBstMidEndSepPunct{\mcitedefaultmidpunct}
{}{\mcitedefaultseppunct}\relax
\EndOfBibitem
\bibitem[Barker(1965)]{barker1965monte}
Barker,~A.~A. Monte carlo calculations of the radial distribution functions for
  a proton-electron plasma. \emph{Aust. J. Phys.} \textbf{1965}, \emph{18},
  119--134\relax
\mciteBstWouldAddEndPuncttrue
\mciteSetBstMidEndSepPunct{\mcitedefaultmidpunct}
{\mcitedefaultendpunct}{\mcitedefaultseppunct}\relax
\EndOfBibitem
\bibitem[Liu and Liu(2001)Liu, and Liu]{liu2001monte}
Liu,~J.~S.; Liu,~J.~S. \emph{Monte Carlo strategies in scientific computing};
  Springer, 2001; Vol.~75\relax
\mciteBstWouldAddEndPuncttrue
\mciteSetBstMidEndSepPunct{\mcitedefaultmidpunct}
{\mcitedefaultendpunct}{\mcitedefaultseppunct}\relax
\EndOfBibitem
\bibitem[Liu(1996)]{liu1996peskun}
Liu,~J.~S. Peskun's theorem and a modified discrete-state Gibbs sampler.
  \emph{Biometrika} \textbf{1996}, \emph{83}\relax
\mciteBstWouldAddEndPuncttrue
\mciteSetBstMidEndSepPunct{\mcitedefaultmidpunct}
{\mcitedefaultendpunct}{\mcitedefaultseppunct}\relax
\EndOfBibitem
\bibitem[Marinari and Parisi(1992)Marinari, and Parisi]{marinari1992simulated}
Marinari,~E.; Parisi,~G. Simulated tempering: a new Monte Carlo scheme.
  \emph{EPL} \textbf{1992}, \emph{19}, 451\relax
\mciteBstWouldAddEndPuncttrue
\mciteSetBstMidEndSepPunct{\mcitedefaultmidpunct}
{\mcitedefaultendpunct}{\mcitedefaultseppunct}\relax
\EndOfBibitem
\bibitem[Sugita and Okamoto(1999)Sugita, and Okamoto]{TREMD}
Sugita,~Y.; Okamoto,~Y. Replica-exchange molecular dynamics method for protein
  folding. \emph{Chem. Phys. Lett.} \textbf{1999}, \emph{314}, 141--151\relax
\mciteBstWouldAddEndPuncttrue
\mciteSetBstMidEndSepPunct{\mcitedefaultmidpunct}
{\mcitedefaultendpunct}{\mcitedefaultseppunct}\relax
\EndOfBibitem
\bibitem[Paschek and Garc{\'\i}a(2004)Paschek, and
  Garc{\'\i}a]{paschek2004reversible}
Paschek,~D.; Garc{\'\i}a,~A.~E. Reversible temperature and pressure
  denaturation of a protein fragment: a replica exchange molecular dynamics
  simulation study. \emph{Physical review letters} \textbf{2004}, \emph{93},
  238105\relax
\mciteBstWouldAddEndPuncttrue
\mciteSetBstMidEndSepPunct{\mcitedefaultmidpunct}
{\mcitedefaultendpunct}{\mcitedefaultseppunct}\relax
\EndOfBibitem
\bibitem[Zhou(2006)]{zhou2006replica}
Zhou,~R. Replica exchange molecular dynamics method for protein folding
  simulation. \emph{Protein Folding Protocols} \textbf{2006}, 205--223\relax
\mciteBstWouldAddEndPuncttrue
\mciteSetBstMidEndSepPunct{\mcitedefaultmidpunct}
{\mcitedefaultendpunct}{\mcitedefaultseppunct}\relax
\EndOfBibitem
\bibitem[Li \latin{et~al.}(2007)Li, Fajer, and Yang]{li2007simulated}
Li,~H.; Fajer,~M.; Yang,~W. Simulated scaling method for localized enhanced
  sampling and simultaneous “alchemical” free energy simulations: A general
  method for molecular mechanical, quantum mechanical, and quantum
  mechanical/molecular mechanical simulations. \emph{The Journal of chemical
  physics} \textbf{2007}, \emph{126}, 01B606\relax
\mciteBstWouldAddEndPuncttrue
\mciteSetBstMidEndSepPunct{\mcitedefaultmidpunct}
{\mcitedefaultendpunct}{\mcitedefaultseppunct}\relax
\EndOfBibitem
\bibitem[Lyubartsev \latin{et~al.}(1992)Lyubartsev, Martsinovski, Shevkunov,
  and Vorontsov-Velyaminov]{EXE}
Lyubartsev,~A.; Martsinovski,~A.; Shevkunov,~S.; Vorontsov-Velyaminov,~P. New
  approach to Monte Carlo calculation of the free energy: Method of expanded
  ensembles. \emph{J. Chem. Phys.} \textbf{1992}, \emph{96}, 1776--1783\relax
\mciteBstWouldAddEndPuncttrue
\mciteSetBstMidEndSepPunct{\mcitedefaultmidpunct}
{\mcitedefaultendpunct}{\mcitedefaultseppunct}\relax
\EndOfBibitem
\bibitem[Knight and Brooks~III(2009)Knight, and Brooks~III]{knight2009lambda}
Knight,~J.~L.; Brooks~III,~C.~L. $\lambda$-Dynamics free energy simulation
  methods. \emph{J. Comput. Chem.} \textbf{2009}, \emph{30}, 1692--1700\relax
\mciteBstWouldAddEndPuncttrue
\mciteSetBstMidEndSepPunct{\mcitedefaultmidpunct}
{\mcitedefaultendpunct}{\mcitedefaultseppunct}\relax
\EndOfBibitem
\bibitem[Knight and Brooks~III(2011)Knight, and
  Brooks~III]{knight2011multisite}
Knight,~J.~L.; Brooks~III,~C.~L. Multisite $\lambda$ dynamics for simulated
  structure--activity relationship studies. \emph{Journal of chemical theory
  and computation} \textbf{2011}, \emph{7}, 2728--2739\relax
\mciteBstWouldAddEndPuncttrue
\mciteSetBstMidEndSepPunct{\mcitedefaultmidpunct}
{\mcitedefaultendpunct}{\mcitedefaultseppunct}\relax
\EndOfBibitem
\bibitem[Sugita \latin{et~al.}(2000)Sugita, Kitao, and Okamoto]{HREMD}
Sugita,~Y.; Kitao,~A.; Okamoto,~Y. Multidimensional replica-exchange method for
  free-energy calculations. \emph{J. Chem. Phys.} \textbf{2000}, \emph{113},
  6042--6051\relax
\mciteBstWouldAddEndPuncttrue
\mciteSetBstMidEndSepPunct{\mcitedefaultmidpunct}
{\mcitedefaultendpunct}{\mcitedefaultseppunct}\relax
\EndOfBibitem
\bibitem[Mobley and Guthrie(2014)Mobley, and Guthrie]{mobley2014freesolv}
Mobley,~D.~L.; Guthrie,~J.~P. FreeSolv: a database of experimental and
  calculated hydration free energies, with input files. \emph{Journal of
  computer-aided molecular design} \textbf{2014}, \emph{28}, 711--720\relax
\mciteBstWouldAddEndPuncttrue
\mciteSetBstMidEndSepPunct{\mcitedefaultmidpunct}
{\mcitedefaultendpunct}{\mcitedefaultseppunct}\relax
\EndOfBibitem
\bibitem[Mobley \latin{et~al.}(2012)Mobley, Liu, Cerutti, Swope, and
  Rice]{mobley2012alchemical}
Mobley,~D.~L.; Liu,~S.; Cerutti,~D.~S.; Swope,~W.~C.; Rice,~J.~E. Alchemical
  prediction of hydration free energies for SAMPL. \emph{Journal of
  computer-aided molecular design} \textbf{2012}, \emph{26}, 551--562\relax
\mciteBstWouldAddEndPuncttrue
\mciteSetBstMidEndSepPunct{\mcitedefaultmidpunct}
{\mcitedefaultendpunct}{\mcitedefaultseppunct}\relax
\EndOfBibitem
\bibitem[Scheen \latin{et~al.}(2020)Scheen, Wu, Mey, Tosco, Mackey, and
  Michel]{scheen2020hybrid}
Scheen,~J.; Wu,~W.; Mey,~A.~S.; Tosco,~P.; Mackey,~M.; Michel,~J. Hybrid
  alchemical free energy/machine-learning methodology for the computation of
  hydration free energies. \emph{Journal of Chemical Information and Modeling}
  \textbf{2020}, \emph{60}, 5331--5339\relax
\mciteBstWouldAddEndPuncttrue
\mciteSetBstMidEndSepPunct{\mcitedefaultmidpunct}
{\mcitedefaultendpunct}{\mcitedefaultseppunct}\relax
\EndOfBibitem
\bibitem[Khalak \latin{et~al.}(2021)Khalak, Tresadern, Aldeghi, Baumann,
  Mobley, de~Groot, and Gapsys]{khalak2021alchemical}
Khalak,~Y.; Tresadern,~G.; Aldeghi,~M.; Baumann,~H.~M.; Mobley,~D.~L.;
  de~Groot,~B.~L.; Gapsys,~V. Alchemical absolute protein--ligand binding free
  energies for drug design. \emph{Chemical science} \textbf{2021}, \emph{12},
  13958--13971\relax
\mciteBstWouldAddEndPuncttrue
\mciteSetBstMidEndSepPunct{\mcitedefaultmidpunct}
{\mcitedefaultendpunct}{\mcitedefaultseppunct}\relax
\EndOfBibitem
\bibitem[Lee \latin{et~al.}(2020)Lee, Allen, Giese, Guo, Li, Lin, McGee~Jr,
  Pearlman, Radak, Tao, \latin{et~al.} others]{lee2020alchemical}
Lee,~T.-S.; Allen,~B.~K.; Giese,~T.~J.; Guo,~Z.; Li,~P.; Lin,~C.;
  McGee~Jr,~T.~D.; Pearlman,~D.~A.; Radak,~B.~K.; Tao,~Y., \latin{et~al.}
  Alchemical binding free energy calculations in AMBER20: Advances and best
  practices for drug discovery. \emph{Journal of Chemical Information and
  Modeling} \textbf{2020}, \emph{60}, 5595--5623\relax
\mciteBstWouldAddEndPuncttrue
\mciteSetBstMidEndSepPunct{\mcitedefaultmidpunct}
{\mcitedefaultendpunct}{\mcitedefaultseppunct}\relax
\EndOfBibitem
\bibitem[Abel \latin{et~al.}(2017)Abel, Wang, Mobley, and
  Friesner]{abel2017critical}
Abel,~R.; Wang,~L.; Mobley,~D.~L.; Friesner,~R.~A. A critical review of
  validation, blind testing, and real-world use of alchemical protein-ligand
  binding free energy calculations. \emph{Current topics in medicinal
  chemistry} \textbf{2017}, \emph{17}, 2577--2585\relax
\mciteBstWouldAddEndPuncttrue
\mciteSetBstMidEndSepPunct{\mcitedefaultmidpunct}
{\mcitedefaultendpunct}{\mcitedefaultseppunct}\relax
\EndOfBibitem
\bibitem[Karrenbrock \latin{et~al.}(2023)Karrenbrock, Procacci, and
  Gervasio]{karrenbrock2023nonequilibrium}
Karrenbrock,~M.; Procacci,~P.; Gervasio,~F.~L. A nonequilibrium alchemical
  method for drug-receptor absolute binding free energy calculations: the role
  of restraints. \textbf{2023}, \relax
\mciteBstWouldAddEndPunctfalse
\mciteSetBstMidEndSepPunct{\mcitedefaultmidpunct}
{}{\mcitedefaultseppunct}\relax
\EndOfBibitem
\bibitem[Piomponi \latin{et~al.}(2022)Piomponi, Fr\"{o}hlking, Bernetti, and
  Bussi]{piomponi2022molecular}
Piomponi,~V.; Fr\"{o}hlking,~T.; Bernetti,~M.; Bussi,~G. Molecular simulations
  matching denaturation experiments for N6-Methyladenosine. \emph{ACS Cent.
  Sci.} \textbf{2022}, \emph{8}, 1218--1228\relax
\mciteBstWouldAddEndPuncttrue
\mciteSetBstMidEndSepPunct{\mcitedefaultmidpunct}
{\mcitedefaultendpunct}{\mcitedefaultseppunct}\relax
\EndOfBibitem
\bibitem[Hayes and Brooks~III(2021)Hayes, and Brooks~III]{hayes2021strategy}
Hayes,~R.~L.; Brooks~III,~C.~L. A strategy for proline and glycine mutations to
  proteins with alchemical free energy calculations. \emph{Journal of
  computational chemistry} \textbf{2021}, \emph{42}, 1088--1094\relax
\mciteBstWouldAddEndPuncttrue
\mciteSetBstMidEndSepPunct{\mcitedefaultmidpunct}
{\mcitedefaultendpunct}{\mcitedefaultseppunct}\relax
\EndOfBibitem
\bibitem[Bergonzo \latin{et~al.}(2014)Bergonzo, Henriksen, Roe, Swails,
  Roitberg, and Cheatham~III]{bergonzo2014multidimensional}
Bergonzo,~C.; Henriksen,~N.~M.; Roe,~D.~R.; Swails,~J.~M.; Roitberg,~A.~E.;
  Cheatham~III,~T.~E. Multidimensional replica exchange molecular dynamics
  yields a converged ensemble of an RNA tetranucleotide. \emph{Journal of
  Chemical Theory and Computation} \textbf{2014}, \emph{10}, 492--499\relax
\mciteBstWouldAddEndPuncttrue
\mciteSetBstMidEndSepPunct{\mcitedefaultmidpunct}
{\mcitedefaultendpunct}{\mcitedefaultseppunct}\relax
\EndOfBibitem
\bibitem[Jiang and Roux(2010)Jiang, and Roux]{jiang2010free}
Jiang,~W.; Roux,~B. Free energy perturbation Hamiltonian replica-exchange
  molecular dynamics (FEP/H-REMD) for absolute ligand binding free energy
  calculations. \emph{Journal of chemical theory and computation}
  \textbf{2010}, \emph{6}, 2559--2565\relax
\mciteBstWouldAddEndPuncttrue
\mciteSetBstMidEndSepPunct{\mcitedefaultmidpunct}
{\mcitedefaultendpunct}{\mcitedefaultseppunct}\relax
\EndOfBibitem
\bibitem[Ebrahimi \latin{et~al.}(2019)Ebrahimi, Kaur, Baronti, Petzold, and
  Chen]{ebrahimi2019two}
Ebrahimi,~P.; Kaur,~S.; Baronti,~L.; Petzold,~K.; Chen,~A.~A. A two-dimensional
  replica-exchange molecular dynamics method for simulating RNA folding using
  sparse experimental restraints. \emph{Methods} \textbf{2019}, \emph{162},
  96--107\relax
\mciteBstWouldAddEndPuncttrue
\mciteSetBstMidEndSepPunct{\mcitedefaultmidpunct}
{\mcitedefaultendpunct}{\mcitedefaultseppunct}\relax
\EndOfBibitem
\bibitem[Cruzeiro and Roitberg(2019)Cruzeiro, and
  Roitberg]{cruzeiro2019multidimensional}
Cruzeiro,~V. W.~D.; Roitberg,~A.~E. Multidimensional replica exchange
  simulations for efficient constant pH and redox potential molecular dynamics.
  \emph{Journal of Chemical Theory and Computation} \textbf{2019}, \emph{15},
  871--881\relax
\mciteBstWouldAddEndPuncttrue
\mciteSetBstMidEndSepPunct{\mcitedefaultmidpunct}
{\mcitedefaultendpunct}{\mcitedefaultseppunct}\relax
\EndOfBibitem
\bibitem[Wang and Landau(2001)Wang, and Landau]{wang2001efficient}
Wang,~F.; Landau,~D.~P. Efficient, multiple-range random walk algorithm to
  calculate the density of states. \emph{Phys. Rev. Lett.} \textbf{2001},
  \emph{86}, 2050\relax
\mciteBstWouldAddEndPuncttrue
\mciteSetBstMidEndSepPunct{\mcitedefaultmidpunct}
{\mcitedefaultendpunct}{\mcitedefaultseppunct}\relax
\EndOfBibitem
\bibitem[Belardinelli and Pereyra(2007)Belardinelli, and
  Pereyra]{belardinelli2007fast}
Belardinelli,~R.; Pereyra,~V. Fast algorithm to calculate density of states.
  \emph{Phys. Rev. E} \textbf{2007}, \emph{75}, 046701\relax
\mciteBstWouldAddEndPuncttrue
\mciteSetBstMidEndSepPunct{\mcitedefaultmidpunct}
{\mcitedefaultendpunct}{\mcitedefaultseppunct}\relax
\EndOfBibitem
\bibitem[Belardinelli and Pereyra(2007)Belardinelli, and
  Pereyra]{belardinelli2007wang}
Belardinelli,~R.; Pereyra,~V. Wang-Landau algorithm: A theoretical analysis of
  the saturation of the error. \emph{The Journal of chemical physics}
  \textbf{2007}, \emph{127}, 184105\relax
\mciteBstWouldAddEndPuncttrue
\mciteSetBstMidEndSepPunct{\mcitedefaultmidpunct}
{\mcitedefaultendpunct}{\mcitedefaultseppunct}\relax
\EndOfBibitem
\bibitem[Lidmar(2012)]{lidmar2012improving}
Lidmar,~J. Improving the efficiency of extended ensemble simulations: The
  accelerated weight histogram method. \emph{Physical Review E} \textbf{2012},
  \emph{85}, 056708\relax
\mciteBstWouldAddEndPuncttrue
\mciteSetBstMidEndSepPunct{\mcitedefaultmidpunct}
{\mcitedefaultendpunct}{\mcitedefaultseppunct}\relax
\EndOfBibitem
\bibitem[Lindahl \latin{et~al.}(2014)Lindahl, Lidmar, and
  Hess]{lindahl2014accelerated}
Lindahl,~V.; Lidmar,~J.; Hess,~B. Accelerated weight histogram method for
  exploring free energy landscapes. \emph{The Journal of chemical physics}
  \textbf{2014}, \emph{141}\relax
\mciteBstWouldAddEndPuncttrue
\mciteSetBstMidEndSepPunct{\mcitedefaultmidpunct}
{\mcitedefaultendpunct}{\mcitedefaultseppunct}\relax
\EndOfBibitem
\bibitem[Lundborg \latin{et~al.}(2021)Lundborg, Lidmar, and
  Hess]{lundborg2021accelerated}
Lundborg,~M.; Lidmar,~J.; Hess,~B. The accelerated weight histogram method for
  alchemical free energy calculations. \emph{The Journal of Chemical Physics}
  \textbf{2021}, \emph{154}\relax
\mciteBstWouldAddEndPuncttrue
\mciteSetBstMidEndSepPunct{\mcitedefaultmidpunct}
{\mcitedefaultendpunct}{\mcitedefaultseppunct}\relax
\EndOfBibitem
\bibitem[Tan(2017)]{tan2017optimally}
Tan,~Z. Optimally adjusted mixture sampling and locally weighted histogram
  analysis. \emph{Journal of Computational and Graphical Statistics}
  \textbf{2017}, \emph{26}, 54--65\relax
\mciteBstWouldAddEndPuncttrue
\mciteSetBstMidEndSepPunct{\mcitedefaultmidpunct}
{\mcitedefaultendpunct}{\mcitedefaultseppunct}\relax
\EndOfBibitem
\bibitem[Hsu \latin{et~al.}(2023)Hsu, Piomponi, Merz, Bussi, and
  Shirts]{hsu2023alchemical}
Hsu,~W.-T.; Piomponi,~V.; Merz,~P.~T.; Bussi,~G.; Shirts,~M.~R. Alchemical
  Metadynamics: Adding Alchemical Variables to Metadynamics to Enhance Sampling
  in Free Energy Calculations. \emph{Journal of Chemical Theory and
  Computation} \textbf{2023}, \emph{19}, 1805--1817\relax
\mciteBstWouldAddEndPuncttrue
\mciteSetBstMidEndSepPunct{\mcitedefaultmidpunct}
{\mcitedefaultendpunct}{\mcitedefaultseppunct}\relax
\EndOfBibitem
\bibitem[Kirkwood(1935)]{kirkwood1935statistical}
Kirkwood,~J.~G. Statistical mechanics of fluid mixtures. \emph{The Journal of
  chemical physics} \textbf{1935}, \emph{3}, 300--313\relax
\mciteBstWouldAddEndPuncttrue
\mciteSetBstMidEndSepPunct{\mcitedefaultmidpunct}
{\mcitedefaultendpunct}{\mcitedefaultseppunct}\relax
\EndOfBibitem
\bibitem[Bennett(1976)]{bennett1976efficient}
Bennett,~C.~H. Efficient estimation of free energy differences from Monte Carlo
  data. \emph{Journal of Computational Physics} \textbf{1976}, \emph{22},
  245--268\relax
\mciteBstWouldAddEndPuncttrue
\mciteSetBstMidEndSepPunct{\mcitedefaultmidpunct}
{\mcitedefaultendpunct}{\mcitedefaultseppunct}\relax
\EndOfBibitem
\bibitem[Shirts and Chodera(2008)Shirts, and Chodera]{shirts2008statistically}
Shirts,~M.~R.; Chodera,~J.~D. Statistically optimal analysis of samples from
  multiple equilibrium states. \emph{The Journal of chemical physics}
  \textbf{2008}, \emph{129}, 124105\relax
\mciteBstWouldAddEndPuncttrue
\mciteSetBstMidEndSepPunct{\mcitedefaultmidpunct}
{\mcitedefaultendpunct}{\mcitedefaultseppunct}\relax
\EndOfBibitem
\bibitem[Escobedo and Martinez-Veracoechea(2008)Escobedo, and
  Martinez-Veracoechea]{escobedo2008optimization}
Escobedo,~F.~A.; Martinez-Veracoechea,~F.~J. Optimization of expanded ensemble
  methods. \emph{The Journal of chemical physics} \textbf{2008},
  \emph{129}\relax
\mciteBstWouldAddEndPuncttrue
\mciteSetBstMidEndSepPunct{\mcitedefaultmidpunct}
{\mcitedefaultendpunct}{\mcitedefaultseppunct}\relax
\EndOfBibitem
\bibitem[Monroe and Shirts(2014)Monroe, and Shirts]{monroe2014converging}
Monroe,~J.~I.; Shirts,~M.~R. Converging free energies of binding in cucurbit
  [7] uril and octa-acid host--guest systems from SAMPL4 using expanded
  ensemble simulations. \emph{Journal of computer-aided molecular design}
  \textbf{2014}, \emph{28}, 401--415\relax
\mciteBstWouldAddEndPuncttrue
\mciteSetBstMidEndSepPunct{\mcitedefaultmidpunct}
{\mcitedefaultendpunct}{\mcitedefaultseppunct}\relax
\EndOfBibitem
\bibitem[Rizzi \latin{et~al.}(2020)Rizzi, Jensen, Slochower, Aldeghi, Gapsys,
  Ntekoumes, Bosisio, Papadourakis, Henriksen, De~Groot, \latin{et~al.}
  others]{rizzi2020sampl6}
Rizzi,~A.; Jensen,~T.; Slochower,~D.~R.; Aldeghi,~M.; Gapsys,~V.;
  Ntekoumes,~D.; Bosisio,~S.; Papadourakis,~M.; Henriksen,~N.~M.;
  De~Groot,~B.~L., \latin{et~al.}  The SAMPL6 SAMPLing challenge: assessing the
  reliability and efficiency of binding free energy calculations. \emph{Journal
  of computer-aided molecular design} \textbf{2020}, \emph{34}, 601--633\relax
\mciteBstWouldAddEndPuncttrue
\mciteSetBstMidEndSepPunct{\mcitedefaultmidpunct}
{\mcitedefaultendpunct}{\mcitedefaultseppunct}\relax
\EndOfBibitem
\bibitem[Muddana \latin{et~al.}(2014)Muddana, Fenley, Mobley, and
  Gilson]{muddana2014sampl4}
Muddana,~H.~S.; Fenley,~A.~T.; Mobley,~D.~L.; Gilson,~M.~K. The SAMPL4
  host--guest blind prediction challenge: an overview. \emph{Journal of
  computer-aided molecular design} \textbf{2014}, \emph{28}, 305--317\relax
\mciteBstWouldAddEndPuncttrue
\mciteSetBstMidEndSepPunct{\mcitedefaultmidpunct}
{\mcitedefaultendpunct}{\mcitedefaultseppunct}\relax
\EndOfBibitem
\bibitem[Gallicchio \latin{et~al.}(2008)Gallicchio, Levy, and
  Parashar]{gallicchio2008asynchronous}
Gallicchio,~E.; Levy,~R.~M.; Parashar,~M. Asynchronous replica exchange for
  molecular simulations. \emph{Journal of computational chemistry}
  \textbf{2008}, \emph{29}, 788--794\relax
\mciteBstWouldAddEndPuncttrue
\mciteSetBstMidEndSepPunct{\mcitedefaultmidpunct}
{\mcitedefaultendpunct}{\mcitedefaultseppunct}\relax
\EndOfBibitem
\bibitem[Gallicchio \latin{et~al.}(2015)Gallicchio, Xia, Flynn, Zhang,
  Samlalsingh, Mentes, and Levy]{gallicchio2015asynchronous}
Gallicchio,~E.; Xia,~J.; Flynn,~W.~F.; Zhang,~B.; Samlalsingh,~S.; Mentes,~A.;
  Levy,~R.~M. Asynchronous replica exchange software for grid and heterogeneous
  computing. \emph{Computer physics communications} \textbf{2015}, \emph{196},
  236--246\relax
\mciteBstWouldAddEndPuncttrue
\mciteSetBstMidEndSepPunct{\mcitedefaultmidpunct}
{\mcitedefaultendpunct}{\mcitedefaultseppunct}\relax
\EndOfBibitem
\bibitem[Xia \latin{et~al.}(2015)Xia, Flynn, Gallicchio, Zhang, He, Tan, and
  Levy]{xia2015large}
Xia,~J.; Flynn,~W.~F.; Gallicchio,~E.; Zhang,~B.~W.; He,~P.; Tan,~Z.;
  Levy,~R.~M. Large-scale asynchronous and distributed multidimensional replica
  exchange molecular simulations and efficiency analysis. \emph{Journal of
  computational chemistry} \textbf{2015}, \emph{36}, 1772--1785\relax
\mciteBstWouldAddEndPuncttrue
\mciteSetBstMidEndSepPunct{\mcitedefaultmidpunct}
{\mcitedefaultendpunct}{\mcitedefaultseppunct}\relax
\EndOfBibitem
\bibitem[Radak \latin{et~al.}(2013)Radak, Romanus, Gallicchio, Lee, Weidner,
  Deng, He, Dai, York, Levy, \latin{et~al.} others]{radak2013framework}
Radak,~B.~K.; Romanus,~M.; Gallicchio,~E.; Lee,~T.-S.; Weidner,~O.;
  Deng,~N.-J.; He,~P.; Dai,~W.; York,~D.~M.; Levy,~R.~M., \latin{et~al.}  A
  framework for flexible and scalable replica-exchange on production
  distributed CI. Proceedings of the Conference on Extreme Science and
  Engineering Discovery Environment: Gateway to Discovery. 2013; pp 1--8\relax
\mciteBstWouldAddEndPuncttrue
\mciteSetBstMidEndSepPunct{\mcitedefaultmidpunct}
{\mcitedefaultendpunct}{\mcitedefaultseppunct}\relax
\EndOfBibitem
\bibitem[Lockhart \latin{et~al.}(2015)Lockhart, O’Connor, Armentrout, and
  Klimov]{lockhart2015greedy}
Lockhart,~C.; O’Connor,~J.; Armentrout,~S.; Klimov,~D.~K. Greedy replica
  exchange algorithm for heterogeneous computing grids. \emph{Journal of
  molecular modeling} \textbf{2015}, \emph{21}, 1--12\relax
\mciteBstWouldAddEndPuncttrue
\mciteSetBstMidEndSepPunct{\mcitedefaultmidpunct}
{\mcitedefaultendpunct}{\mcitedefaultseppunct}\relax
\EndOfBibitem
\bibitem[Rhee and Pande(2003)Rhee, and Pande]{rhee2003multiplexed}
Rhee,~Y.~M.; Pande,~V.~S. Multiplexed-replica exchange molecular dynamics
  method for protein folding simulation. \emph{Biophysical journal}
  \textbf{2003}, \emph{84}, 775--786\relax
\mciteBstWouldAddEndPuncttrue
\mciteSetBstMidEndSepPunct{\mcitedefaultmidpunct}
{\mcitedefaultendpunct}{\mcitedefaultseppunct}\relax
\EndOfBibitem
\bibitem[Paliwal and Shirts(2011)Paliwal, and Shirts]{paliwal2011benchmark}
Paliwal,~H.; Shirts,~M.~R. A benchmark test set for alchemical free energy
  transformations and its use to quantify error in common free energy methods.
  \emph{Journal of chemical theory and computation} \textbf{2011}, \emph{7},
  4115--4134\relax
\mciteBstWouldAddEndPuncttrue
\mciteSetBstMidEndSepPunct{\mcitedefaultmidpunct}
{\mcitedefaultendpunct}{\mcitedefaultseppunct}\relax
\EndOfBibitem
\bibitem[Hess \latin{et~al.}(2008)Hess, Kutzner, Van Der~Spoel, and
  Lindahl]{hess2008gromacs}
Hess,~B.; Kutzner,~C.; Van Der~Spoel,~D.; Lindahl,~E. GROMACS 4: algorithms for
  highly efficient, load-balanced, and scalable molecular simulation.
  \emph{Journal of chemical theory and computation} \textbf{2008}, \emph{4},
  435--447\relax
\mciteBstWouldAddEndPuncttrue
\mciteSetBstMidEndSepPunct{\mcitedefaultmidpunct}
{\mcitedefaultendpunct}{\mcitedefaultseppunct}\relax
\EndOfBibitem
\bibitem[Pronk \latin{et~al.}(2013)Pronk, P{\'a}ll, Schulz, Larsson, Bjelkmar,
  Apostolov, Shirts, Smith, Kasson, Van Der~Spoel, \latin{et~al.}
  others]{pronk2013gromacs}
Pronk,~S.; P{\'a}ll,~S.; Schulz,~R.; Larsson,~P.; Bjelkmar,~P.; Apostolov,~R.;
  Shirts,~M.~R.; Smith,~J.~C.; Kasson,~P.~M.; Van Der~Spoel,~D., \latin{et~al.}
   GROMACS 4.5: a high-throughput and highly parallel open source molecular
  simulation toolkit. \emph{Bioinformatics} \textbf{2013}, \emph{29},
  845--854\relax
\mciteBstWouldAddEndPuncttrue
\mciteSetBstMidEndSepPunct{\mcitedefaultmidpunct}
{\mcitedefaultendpunct}{\mcitedefaultseppunct}\relax
\EndOfBibitem
\bibitem[Chodera(2016)]{chodera2016simple}
Chodera,~J.~D. A simple method for automated equilibration detection in
  molecular simulations. \emph{Journal of chemical theory and computation}
  \textbf{2016}, \emph{12}, 1799--1805\relax
\mciteBstWouldAddEndPuncttrue
\mciteSetBstMidEndSepPunct{\mcitedefaultmidpunct}
{\mcitedefaultendpunct}{\mcitedefaultseppunct}\relax
\EndOfBibitem
\bibitem[Husic and Pande(2018)Husic, and Pande]{husic2018markov}
Husic,~B.~E.; Pande,~V.~S. Markov state models: From an art to a science.
  \emph{Journal of the American Chemical Society} \textbf{2018}, \emph{140},
  2386--2396\relax
\mciteBstWouldAddEndPuncttrue
\mciteSetBstMidEndSepPunct{\mcitedefaultmidpunct}
{\mcitedefaultendpunct}{\mcitedefaultseppunct}\relax
\EndOfBibitem
\bibitem[Bowman \latin{et~al.}(2013)Bowman, Pande, and
  No{\'e}]{bowman2013introduction}
Bowman,~G.~R.; Pande,~V.~S.; No{\'e},~F. \emph{An introduction to Markov state
  models and their application to long timescale molecular simulation};
  Springer Science \& Business Media, 2013; Vol. 797\relax
\mciteBstWouldAddEndPuncttrue
\mciteSetBstMidEndSepPunct{\mcitedefaultmidpunct}
{\mcitedefaultendpunct}{\mcitedefaultseppunct}\relax
\EndOfBibitem
\bibitem[Scherer \latin{et~al.}(2015)Scherer, Trendelkamp-Schroer, Paul,
  P{\'e}rez-Hern{\'a}ndez, Hoffmann, Plattner, Wehmeyer, Prinz, and
  No{\'e}]{scherer2015pyemma}
Scherer,~M.~K.; Trendelkamp-Schroer,~B.; Paul,~F.; P{\'e}rez-Hern{\'a}ndez,~G.;
  Hoffmann,~M.; Plattner,~N.; Wehmeyer,~C.; Prinz,~J.-H.; No{\'e},~F. PyEMMA 2:
  A software package for estimation, validation, and analysis of Markov models.
  \emph{Journal of chemical theory and computation} \textbf{2015}, \emph{11},
  5525--5542\relax
\mciteBstWouldAddEndPuncttrue
\mciteSetBstMidEndSepPunct{\mcitedefaultmidpunct}
{\mcitedefaultendpunct}{\mcitedefaultseppunct}\relax
\EndOfBibitem
\bibitem[Bussi \latin{et~al.}(2007)Bussi, Donadio, and
  Parrinello]{bussi2007canonical}
Bussi,~G.; Donadio,~D.; Parrinello,~M. Canonical sampling through velocity
  rescaling. \emph{J. Chem. Phys.} \textbf{2007}, \emph{126}, 014101\relax
\mciteBstWouldAddEndPuncttrue
\mciteSetBstMidEndSepPunct{\mcitedefaultmidpunct}
{\mcitedefaultendpunct}{\mcitedefaultseppunct}\relax
\EndOfBibitem
\bibitem[Berendsen \latin{et~al.}(1984)Berendsen, Postma, Van~Gunsteren,
  DiNola, and Haak]{berendsen1984molecular}
Berendsen,~H.~J.; Postma,~J.~v.; Van~Gunsteren,~W.~F.; DiNola,~A.; Haak,~J.~R.
  Molecular dynamics with coupling to an external bath. \emph{J. Chem. Phys.}
  \textbf{1984}, \emph{81}, 3684--3690\relax
\mciteBstWouldAddEndPuncttrue
\mciteSetBstMidEndSepPunct{\mcitedefaultmidpunct}
{\mcitedefaultendpunct}{\mcitedefaultseppunct}\relax
\EndOfBibitem
\bibitem[Parrinello and Rahman(1980)Parrinello, and
  Rahman]{parrinello1980crystal}
Parrinello,~M.; Rahman,~A. Crystal structure and pair potentials: A
  molecular-dynamics study. \emph{Phys. Rev. Lett.} \textbf{1980}, \emph{45},
  1196\relax
\mciteBstWouldAddEndPuncttrue
\mciteSetBstMidEndSepPunct{\mcitedefaultmidpunct}
{\mcitedefaultendpunct}{\mcitedefaultseppunct}\relax
\EndOfBibitem
\bibitem[Parrinello and Rahman(1981)Parrinello, and
  Rahman]{parrinello1981polymorphic}
Parrinello,~M.; Rahman,~A. Polymorphic transitions in single crystals: A new
  molecular dynamics method. \emph{J. Appl. Phys.} \textbf{1981}, \emph{52},
  7182--7190\relax
\mciteBstWouldAddEndPuncttrue
\mciteSetBstMidEndSepPunct{\mcitedefaultmidpunct}
{\mcitedefaultendpunct}{\mcitedefaultseppunct}\relax
\EndOfBibitem
\bibitem[Essmann \latin{et~al.}(1995)Essmann, Perera, Berkowitz, Darden, Lee,
  and Pedersen]{essmann1995smooth}
Essmann,~U.; Perera,~L.; Berkowitz,~M.~L.; Darden,~T.; Lee,~H.; Pedersen,~L.~G.
  A smooth particle mesh Ewald method. \emph{J. Chem. Phys.} \textbf{1995},
  \emph{103}, 8577--8593\relax
\mciteBstWouldAddEndPuncttrue
\mciteSetBstMidEndSepPunct{\mcitedefaultmidpunct}
{\mcitedefaultendpunct}{\mcitedefaultseppunct}\relax
\EndOfBibitem
\bibitem[Hess \latin{et~al.}(1997)Hess, Bekker, Berendsen, and
  Fraaije]{hess1997lincs}
Hess,~B.; Bekker,~H.; Berendsen,~H.~J.; Fraaije,~J.~G. LINCS: a linear
  constraint solver for molecular simulations. \emph{J. Comput. Chem.}
  \textbf{1997}, \emph{18}, 1463--1472\relax
\mciteBstWouldAddEndPuncttrue
\mciteSetBstMidEndSepPunct{\mcitedefaultmidpunct}
{\mcitedefaultendpunct}{\mcitedefaultseppunct}\relax
\EndOfBibitem
\bibitem[Iorga and Mobley(2021)Iorga, and Mobley]{IorgaMobley2021}
Iorga,~B.~I.; Mobley,~D.~L. samplchallenges/SAMPL4: Version 1.0: Historical
  data. 2021; \url{https://doi.org/10.5281/zenodo.5508284}, Version 1.0:
  Historical data\relax
\mciteBstWouldAddEndPuncttrue
\mciteSetBstMidEndSepPunct{\mcitedefaultmidpunct}
{\mcitedefaultendpunct}{\mcitedefaultseppunct}\relax
\EndOfBibitem
\bibitem[Gilson \latin{et~al.}(1997)Gilson, Given, Bush, and
  McCammon]{gilson1997statistical}
Gilson,~M.~K.; Given,~J.~A.; Bush,~B.~L.; McCammon,~J.~A. The
  statistical-thermodynamic basis for computation of binding affinities: a
  critical review. \emph{Biophysical journal} \textbf{1997}, \emph{72},
  1047--1069\relax
\mciteBstWouldAddEndPuncttrue
\mciteSetBstMidEndSepPunct{\mcitedefaultmidpunct}
{\mcitedefaultendpunct}{\mcitedefaultseppunct}\relax
\EndOfBibitem
\bibitem[Park(2008)]{park2008comparison}
Park,~S. Comparison of the serial and parallel algorithms of generalized
  ensemble simulations: An analytical approach. \emph{Physical Review E}
  \textbf{2008}, \emph{77}, 016709\relax
\mciteBstWouldAddEndPuncttrue
\mciteSetBstMidEndSepPunct{\mcitedefaultmidpunct}
{\mcitedefaultendpunct}{\mcitedefaultseppunct}\relax
\EndOfBibitem
\bibitem[Sindhikara \latin{et~al.}(2008)Sindhikara, Meng, and
  Roitberg]{sindhikara2008exchange}
Sindhikara,~D.; Meng,~Y.; Roitberg,~A.~E. Exchange frequency in replica
  exchange molecular dynamics. \emph{The Journal of chemical physics}
  \textbf{2008}, \emph{128}\relax
\mciteBstWouldAddEndPuncttrue
\mciteSetBstMidEndSepPunct{\mcitedefaultmidpunct}
{\mcitedefaultendpunct}{\mcitedefaultseppunct}\relax
\EndOfBibitem
\bibitem[Sindhikara \latin{et~al.}(2010)Sindhikara, Emerson, and
  Roitberg]{sindhikara2010exchange}
Sindhikara,~D.~J.; Emerson,~D.~J.; Roitberg,~A.~E. Exchange often and properly
  in replica exchange molecular dynamics. \emph{Journal of Chemical Theory and
  Computation} \textbf{2010}, \emph{6}, 2804--2808\relax
\mciteBstWouldAddEndPuncttrue
\mciteSetBstMidEndSepPunct{\mcitedefaultmidpunct}
{\mcitedefaultendpunct}{\mcitedefaultseppunct}\relax
\EndOfBibitem
\end{mcitethebibliography}
